\documentclass[10pt,a4paper]{article}

\usepackage{cite}
\usepackage[latin1]{inputenc}
\usepackage{graphicx}
\usepackage{amsmath}
\usepackage{amsfonts}
\usepackage{amssymb}
\usepackage{makeidx}
\usepackage{setspace}
\usepackage{geometry}
\usepackage{subfigure}
\usepackage{color}
\usepackage{epstopdf}
\usepackage{cite}

\title{Are Performance Limitations in Visual Short-Term Memory Tasks Due to Capacity Limitations or Model Mismatch?}

\author{
A. Emin Orhan \\
E-mail: \texttt{eorhan@cns.nyu.edu} \\
Center for Neural Science, New York University, New York, NY, USA \\ \\
Robert A. Jacobs \\
Department of Brain and Cognitive Sciences, University of Rochester, Rochester, NY, USA 
}

\begin{document}
\maketitle

\section*{Abstract}
Performance limitations in visual short-term memory (VSTM) tasks have traditionally been explained in terms of resource or capacity limitations. It has been claimed, for example, that VSTM possesses a limited amount of cognitive or neural ``resources'' that can be used to remember a visual display. In this paper, we highlight the potential importance of a previously neglected factor that might contribute significantly to performance limitations in VSTM tasks: namely, a mismatch between the prior expectations and/or the internal noise properties of the visual system based primarily on its adaptation to the statistics of the natural environment and the statistics of the visual stimuli used in most VSTM experiments. We call this `model mismatch'. Surprisingly, we show that model mismatch alone, without assuming a general resource or capacity limitation, can, {in principle}, account for some of the main qualitative characteristics of performance limitations observed in VSTM tasks, including: (i) monotonic decline in memory precision with increasing set size; (ii) variability in memory precision across items and trials; and (iii) different set-size dependencies for initial encoding rate and asymptotic precision when the duration of image presentation is varied. We also investigate the consequences of using experimental stimuli that more closely match the prior expectations and/or internal noise properties of the visual system. The results reveal qualitatively very different patterns of behavior for such stimuli, suggesting that researchers should be cautious about generalizing the results of experiments using ecologically unrealistic stimulus statistics to ecologically more realistic stimuli. 

\section*{Introduction}
For a coherent and stable perception of the world, the visual system needs a form of memory that can maintain visual information about the world across brief disruptions and discontinuities in the visual signals reaching the eyes. Such disruptions are quite common in natural vision. For example, visual signals are effectively suppressed during saccadic eye movements or visual signals from a scene are disrupted when an object temporarily occludes the observer's view. It is believed that visual short-term memory (VSTM) is the memory system that subserves the maintenance of visual information across such disruptions \cite{irwin1991}, \cite{irwin1996}, \cite{hollingworthetal2008}. Given the prevalence of such disruptions in natural vision, it is important to understand the properties and limitations of this memory system.

Experimental evidence suggests that surprisingly little information can be maintained in VSTM. In standard VSTM experiments, subjects are briefly shown random visual displays containing simple objects and, after a brief delay interval, their memory for one of the objects is probed using a recall or a recognition task. The precision of memory declines sharply, approximately as a power law, as the number of objects on the display is increased. This is generally attributed to a resource limitation in VSTM. Although the precise nature of the hypothesized ``resources'' is disputed, with different theories favoring either a discrete resource that cannot be divided into arbitrarily small units \cite{zhangluck2008}, \cite{andersonetal2011}, or a more continuous resource \cite{bayshusain2008}, \cite{vandenbergetal2012}, there is little disagreement among researchers that performance limitations in VSTM tasks are caused exclusively by a resource limitation.  

In this paper, we challenge this assumption by demonstrating the potential importance of a previously neglected factor that is independent of resource limitations. We call this factor ``model mismatch''. In general, model mismatch refers to any case where the observer has an imperfect or suboptimal internal model of the stimuli used in an experimental study. Model mismatch causes the observer to perform suboptimal statistical inference and thus leads to suboptimal performance in the experimental task \cite{ma2012}. A mismatch between the true model of the stimuli and the model assumed by the observer can, for example, arise when the observer's internal model is adapted to a type of stimulus statistics that deviates from the stimulus statistics used in the experimental study and the observer has a limited ability to adapt to the actual stimulus statistics during the course of the experiment. 

What type of stimulus statistics could the observer's internal model be adapted to? The hypothesis that will be considered in this paper is that the observer's internal model could be adapted to the statistics of the stimuli in the natural environment. This means that whenever the stimulus statistics used in an experiment deviates from the natural stimulus statistics, as is the case in many VSTM studies, it causes a mismatch between the observer's internal model and the actual stimulus statistics. The severity of the mismatch depends on the difference between the actual stimulus statistics used in the experiment and the natural stimulus statistics as well as on the observer's ability to adapt to the new stimulus statistics. 

Remarkably, we show, through simulations and analytic results, that model mismatch alone can, in principle, account for some of the main qualitative characteristics of performance limitations observed in VSTM tasks without assuming any resource limitation, including: (i) the decline in memory precision with increasing set size \cite{bayshusain2008}, \cite{wilkenma2004}; (ii) variability in memory precision across items and trials \cite{vandenbergetal2012}, \cite{fougnieetal2012}; and (iii) different set-size dependencies for initial encoding rate and asymptotic precision when the duration of image presentation is varied \cite{baysetal2011}. 

Although this result establishes the possibility, in principle, of accounting for performance limitations in VSTM tasks in terms of model mismatch alone, it does not rule out the existence of a resource limitation in VSTM. In reality, both factors might be contributing to performance limitations in VSTM tasks in different degrees. {To do full justice to this possibility, we also investigate the consequences of combining a resource limitation with model mismatch.}

{Finally,} we consider, through simulations, the scenario where the experimenter uses ecologically more realistic stimulus distributions that match the observer's internal model more closely. We show that this leads to qualitatively very different behavioral patterns than the ones obtained with ecologically unrealistic stimulus distributions that do not match the observer's internal model, suggesting that researchers should be cautious about generalizing the results of experiments using ecologically unrealistic stimulus distributions to ecologically more realistic conditions. 

\section*{Methods}

\subsection*{General framework}
\label{general_framework} 

Throughout the paper, we consider a hypothetical VSTM recall task where in each trial a subject is briefly shown a display containing $N$ simple items with feature values $\mathbf{s} = [s_1,\ldots, s_N]$. After a delay interval, one of the items, called the target item, is cued and the subject is asked to recall the feature value of the target. In each trial, the feature values of the items, $\mathbf{s}$, are randomly sampled from a distribution $p(\mathbf{s})$ determined by the experimenter. In standard VSTM experiments, feature values of different items are generally drawn independently from a uniform distribution over a fixed interval. This corresponds to $p(\mathbf{s}) = p(s_1) p(s_2) \cdots p(s_N)$ where each $p(s_i)$ is a uniform distribution over a given range.

The subject is assumed to have access only to noisy measurements or observations, denoted $\mathbf{x} = [x_1,\ldots, x_N]$, of the actual feature values of the items. This observation process can be characterized by the conditional distribution $p(\mathbf{x}|\mathbf{s})$. The optimal strategy for the subject is to combine the actual prior $p(\mathbf{s})$ and the likelihood $p(\mathbf{x}|\mathbf{s})$ to compute the marginal posterior over the feature value of the target, $s_t$, given the noisy observations:
\begin{equation}
p(s_t|\mathbf{x}) \propto \int p(\mathbf{x}|\mathbf{s})p(\mathbf{s}) d\mathbf{s}_{-t}
\label{tru_post}
\end{equation}
where $\mathbf{s}_{-t}$ denotes the feature values of all items except the target item. The subject can generate a point estimate of the target's feature value using this marginal posterior (e.g., use the posterior mean if the mean squared error is to be minimized). However, our framework allows for the possibility that the subject might use a prior, denoted $q(\mathbf{s})$, that is different than the one used by the experimenter to generate the stimuli, thus making responses based on an approximate posterior, $q(s_t|\mathbf{x})$, rather than on the true posterior: 
\begin{equation}
q(s_t|\mathbf{x}) \propto \int p(\mathbf{x}|\mathbf{s})q(\mathbf{s}) d\mathbf{s}_{-t}.
\label{app_post}
\end{equation}
We emphasize that this does not mean that the subject is unaware of the prior used by the experimenter. If, for example, the actual feature values of the items are independently drawn from a uniform distribution, the subject may be aware of this fact. The subject may even be explicitly told by the experimenter how the stimuli are generated. We assume that conscious awareness of stimulus statistics is not a reliable indicator of the particular prior used by the subject in a given experimental task. Rather, the subject's prior $q(\mathbf{s})$ is intended to characterize the visual system's \textit{a priori} expectations about configurations of features (similar to the treatment of prior distributions in \cite{stockersimoncelli2006}, \cite{girshicketal2011}). Below, we motivate our choices for $q(\mathbf{s})$ by some prominent properties of visual features in the natural environment, based on the assumption that the probability of a particular $\mathbf{s}$ under $q(\mathbf{s})$, in general, reflects the statistical regularities of visual features in the natural environment due to the adaptation of the visual system to stimulus statistics in the natural environment. 

In addition to the prior distribution $q(\mathbf{s})$, the noise distribution $p(\mathbf{x}|\mathbf{s})$ may also be adapted to the stimulus statistics in the natural environment. For example, stimuli $\mathbf{s}$ that are more likely in the natural environment may be encoded with better precision, i.e. with less noisy measurements $\mathbf{x}$, under the noise distribution \cite{girshicketal2011}, \cite{wainwright1999}, \cite{gangulisimoncelli2010}. In many standard detection or discrimination tasks, subjects perform better when stimulus features are consistent with natural scene statistics (i.e., when $\mathbf{s}$ is more likely under $q(\mathbf{s})$; see \cite{stockersimoncelli2006}, \cite{girshicketal2011}, \cite{knilletal1990}, \cite{parraghaetal2000}, \cite{yuilleetal2004}). In general, this may reflect the adaptation of both the prior and the noise distribution to natural scene statistics. Girshick et al. \cite{girshicketal2011}, for instance, found that regions in the stimulus space that are more likely under the subject's prior $q(\mathbf{s})$ are also encoded with better precision under the noise distribution. Wei and Stocker \cite{weistocker2012} show that this pattern is consistent with the efficient coding hypothesis \cite{barlow1961} under some plausible assumptions. 

The adaptation of the noise distribution to natural stimulus statistics can be modeled by assuming a stimulus-dependent variance $\sigma^2(\mathbf{s})$ for the noise distribution, with $\mathbf{s}$ that are more likely under $q(\mathbf{s})$ having a smaller noise variance. This possibility will be examined in more detail in a separate subsection under the Results section below. However, elsewhere in the paper we will assume a noise distribution with stimulus-independent covariance, focusing exclusively on the effects of the adaptation of the subject's prior $q(\mathbf{s})$ to stimulus statistics in the natural environment. {\color{black}We will refer to the adaptation of the subject's prior distribution and/or their noise distribution to stimulus statistics different from the actual stimulus statistics used in an experiment as model mismatch.}

\begin{figure}[!ht]
\begin{center}
\includegraphics[scale=0.7]{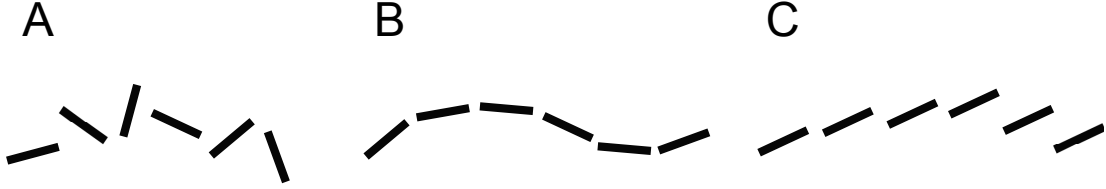} 
\end{center}
\caption{{\bf Illustration of mismatches that can arise between the experimenter's prior $p(\mathbf{s})$ (line orientations are independently sampled from uniform distributions) and the subject's prior $q(\mathbf{s})$ (line orientations are sampled from a joint distribution characterizing orientations in natural visual environments).} The configurations of orientations shown in panels A, B and C have equal probability under $p(\mathbf{s})$, but the configurations in B and C are more probable than the configuration in A under $q(\mathbf{s})$ due to dependencies between the orientations of different line segments in natural images.}
\label{q_illustrate}
\end{figure}

To give a more concrete example of a possible mismatch between the subject's prior and the experimenter's prior, consider the case where $\mathbf{s}$ represents the orientations of items. The experimenter's prior $p(\mathbf{s})$ might be a product of uniform distributions $p(s_i)$, giving all combinations of orientations equal probability. However, all combinations of orientations are not equally likely in the natural world. In fact, only a very small subset of all possible combinations of orientations have significant probability under the statistics of the natural world: for example, parallel orientations or combinations of orientations that form a smooth contour \cite{sigmanetal2001}, \cite{geisleretal2001}. Other combinations of orientations are exceedingly improbable (Figure \ref{q_illustrate}). Through experience-dependent neural plasticity, the visual system is adapted to the statistics of the natural environment. If specific combinations of orientations are highly probable in the natural environment, the visual cortex may allocate more resources to representing those combinations, thereby becoming better at encoding those combinations, whereas random (and thus less likely) combinations of orientations are encoded with greater difficulty or less efficiency.

We do not require that $q(\mathbf{s})$ \textit{exactly} match the statistics of $\mathbf{s}$ in the natural environment. There may be deviations between $q(\mathbf{s})$ and the actual statistics of $\mathbf{s}$ in the natural environment. For example, a particular $\mathbf{s}$ may have a high probability under $q(\mathbf{s})$ for reasons other than being common in the natural environment; for example, accurately detecting that particular $\mathbf{s}$ may be associated with a high reward or survival value. Some researchers have even put forward theoretical arguments against trying to learn an exact model of the environment \cite{feldman2013}. In general, however, $q(\mathbf{s})$ will reflect, more or less faithfully, the statistics of $\mathbf{s}$ in the natural environment, because it is difficult to imagine how an organism would survive with a model of the environment that bears little resemblance to the properties of the actual environment \cite{geislerdiehl2002}. 

Unless explicitly noted otherwise, we employed the following procedure in all simulations reported below. In each simulation, we simulate $10^5$ trials of a hypothetical VSTM recall experiment. In each trial, we first draw $N$ random stimulus values, $\mathbf{s} = [s_1,\ldots, s_N]$, from the experimenter's prior distribution $p(\mathbf{s})$. In different sets of simulations, the experimenter's prior distribution is assumed to be either a multivariate Gaussian distribution with a diagonal covariance matrix and equal marginal variance along each dimension (the standard deviation along each dimension was set to~2 in most simulations) or a multivariate Gaussian with uniform correlations, denoted $\rho_p$, between each pair of its dimensions. In each simulated trial, one of the stimuli, $s_t$, is randomly chosen as the target item. Given the actual values of the stimuli, $\mathbf{s}$, their noisy measurements, $\mathbf{x}$, were generated from the noise distribution $p(\mathbf{x}|\mathbf{s})$. Throughout most of the paper, the noise distribution is assumed to be a multivariate Gaussian with a stimulus-independent diagonal covariance matrix and equal marginal variances along each dimension. In different sets of simulations, the covariance matrix of the noise distribution was either set size independent in which case the standard deviation along each dimension was set to~0.5, or it was dependent on the set size (reflecting a resource limitation) in which case the standard deviation along each dimension was $\lambda_{min} \sqrt{N}$, where $\lambda_{min}=0.5$ and $N$ is the set size. For brevity, only the results for the set size-independent noise distribution are presented in the main text, the results for the set size-dependent noise distribution are discussed briefly. The full set of results for the latter condition can be found in the Supporting Information (Text S5, Figures S5-S8). 

In a separate set of simulations, we also investigate the consequences of adaptation in the noise distribution to natural scene statistics by assuming a stimulus-dependent covariance matrix for $p(\mathbf{x}|\mathbf{s})$ with a smaller noise variance $\sigma^2(\mathbf{s})$ for stimuli that are more likely under $q(\mathbf{s})$, as explained in more detail in the Results section.

Given the noisy measurements, the subject is assumed to compute an approximate marginal posterior over the target stimulus value, using an approximate prior $q(\mathbf{s})$ rather than the true prior $p(\mathbf{s})$ (Equation~\ref{app_post}). We consider different choices for $q(\mathbf{s})$, each motivated by statistical properties of visual features in the natural environment. The mean of the approximate marginal posterior is then taken to be the subject's estimate of the target stimulus value, $\hat{s}_t$, in that trial. For the simulations where the noise distribution is assumed to be adapted to natural scene statistics, the use of a stimulus-dependent noise distribution makes the analytic solution of the joint and marginal posteriors in Equation~\ref{app_post} intractable, thus {we use a Metropolis-Hastings algorithm to sample from the posterior in this case} (Text S1). 

Over many trials, this procedure generates a distribution of errors, $s_t - \hat{s}_t $. The inverse standard deviation ($1/\sigma$) of this error distribution is commonly used as a measure of subjects' performances in VSTM studies. Unfortunately, this quantity is sometimes called recall ``precision'' in the VSTM literature \cite{bayshusain2008}, \cite{baysetal2011}, contrasting with the more standard use of the term ``precision'' in statistics to refer to inverse variance ($1/\sigma^2$). Some VSTM studies use ``precision'' in the sense of inverse variance \cite{vandenbergetal2012}. In what follows, to be consistent with these earlier studies, we use the term ``precision'' in both senses, but we will be explicit about what sense we are intending. We note that although the model responses in individual trials are biased toward the prior mean because of the informative priors we use in this paper, the error distribution itself is not biased (see also Text S3). Additional details about the simulations can be found in the Supporting Information (Text S1).

\subsection*{Models for the subject's prior $q(\mathbf{s})$}  
\label{models_for_q}

We consider three choices for the subject's prior distribution $q(\mathbf{s})$ in our simulations: (i) Gaussian $q(\mathbf{s})$ with uniform, non-negative correlations between stimuli; (ii) Gaussian $q(\mathbf{s})$ with random positive correlations between stimuli; and (iii) a mixture of Gaussians prior favoring homogeneity among the stimuli. These models are all too simple to provide complete characterizations of the subject's prior $q(\mathbf{s})$ presumed to be implemented in the visual cortex, or the relevant environmental statistics of the corresponding visual features. Nonetheless, each contains some crucial properties shared by the environmental statistics of many visual features, hence probably by the prior implemented in the visual cortex as well. Our goal is to demonstrate that important aspects of VSTM performance limitations arise even within the context of these simple models, and thus model mismatch is a viable explanation for these limitations.

\subsubsection*{Gaussian $q(\mathbf{s})$ with uniform non-negative correlations between stimuli} 
\label{q_constcorr}

We first investigate the consequences of a simple mismatch between the correlation structures of the experimenter's and the subject's priors. To this end, we consider the case where the experimenter's prior, $p(\mathbf{s})$, is a multivariate Gaussian distribution with mean $\mathbf{m}$ (we take $\mathbf{m}$ to be a vector of zeros in all simulations), and a covariance matrix with equal marginal variances, $\sigma_p^2$, along each dimension and a uniform non-negative correlation, denoted $\rho_p$, between dimensions (setting $\rho_p=0$ results in a diagonal covariance matrix). The subject's prior $q(\mathbf{s})$ is also assumed to be a multivariate Gaussian with the same mean $\mathbf{m}$ and a covariance matrix with equal marginal variances, $\sigma_q^2$, along each dimension and a uniform non-negative correlation, denoted $\rho_q$, between dimensions. Thus the covariance matrices of both the experimenter's prior and the subject's prior can be written as:
\begin{equation}
\Sigma_N = \begin{bmatrix} \sigma^2 &  & \rho \sigma^2 \\  & \ddots &  \\ \rho \sigma^2 &  & \sigma^2 \end{bmatrix}.
\label{Lambda}
\end{equation}
To investigate the consequences of a mismatch solely between the correlation structures of the experimenter's and the subject's priors, we assume $\sigma_p^2 = \sigma_q^2 = 4$ for the main simulations. However, we also briefly discuss the consequences of an additional mismatch between the variances $\sigma_p^2$ and $\sigma_q^2$. 

Environmental statistics of many basic stimulus features exhibit rich statistical dependencies \cite{sigmanetal2001}, \cite{geisleretal2001}, \cite{hyvarinenetal2009}. Although a Gaussian distribution with covariance matrix given in Equation~\ref{Lambda} is probably not a realistic model of such dependencies for any particular feature dimension, this model is intended to illustrate how even such a simple form of dependency can give rise to prominent performance limitations in VSTM.  

\subsubsection*{Gaussian $q(\mathbf{s})$ with random positive correlations between stimuli} 
\label{q_random}

The second model for the subject's prior $q(\mathbf{s})$ that we consider is a Gaussian distribution with random positive correlations between the stimuli. This can be thought of as a crude approximation to experiments where, in each trial, the stimuli are presented at (pseudo-)random locations on the screen. If it is assumed that correlations between features at different locations decrease as a function of the distance between the locations, as in second-order orientation statistics in natural images \cite{geisleretal2001}, a Gaussian prior with random positive correlations can be considered as an approximate model of the expected correlations between features at random locations. The real natural joint statistics of a number of visual features is more complex than can be captured by a simple Gaussian distribution but, to reiterate, this model is only intended to provide a demonstration of how important VSTM performance limitations can arise even in very simple models. More complex dependencies as exist in the natural statistics of visual features would, if anything, increase the strength of the effects considered in this paper (such as the decline in recall precision with set size) because higher-order dependencies would exacerbate the discrepancy or mismatch between the subject's prior and the experimenter's prior, which usually displays no dependencies.

For simulations of the Gaussian model with random positive correlations, we generated random correlation matrices by first randomly generating eigenvalues for the correlation matrix (Text S1). The eigenvalues are drawn from a symmetric Dirichlet distribution with concentration parameter $\gamma$. Small values of $\gamma$ produce sparse eigenvalues which, in turn, generate correlation matrices with large correlations between dimensions. Larger values of $\gamma$ yield a broader distribution of eigenvalues, which corresponds to smaller correlation values. 

\subsubsection*{Mixture of Gaussians prior favoring homogeneity} 
\label{q_mog}

Natural visual inputs constitute a very small subset of the set of all possible visual inputs. Consider the set of all possible configurations of orientations for $N$ oriented line segments at particular locations in an image. The orientations of these segments can be represented as a vector in a $N$-dimensional space. But naturally-occurring configurations of oriented line segments typically reside in a much smaller dimensional subspace of this $N$-dimensional space of all possible configurations of orientations. Natural configurations will be dominated by relatively homogeneous configurations where many orientations are similar, or by configurations that form smooth contours \cite{geisleretal2001}. If we consider homogeneous configurations where all orientations are similar, these reside near a one-dimensional manifold in the $N$-dimensional space, namely the line $s_1=s_2=\ldots=s_{N}$ where $s_i$ is the orientation of the $i^{\rm th}$ segment. 

Our third model of the subject's prior $q(\mathbf{s})$ is intended to capture the low-dimensional or sparse nature of the natural statistics of visual features. Once again, if we think of the joint statistics of $N$ visual features in a standard VSTM experiment, the experimenter's prior $p(\mathbf{s})$ will typically be a uniform (or at least non-sparse) distribution over the $N$-dimensional space of all possible configurations of the visual features. In contrast, the subject's prior $q(\mathbf{s})$ will have significant probability only over a much lower-dimensional subspace of the $N$-dimensional space. We model this scenario with a mixture of Gaussians prior that places progressively smaller probabilities over higher dimensional subspaces.

\begin{figure}[!ht]
\begin{center}
\includegraphics[scale=0.7]{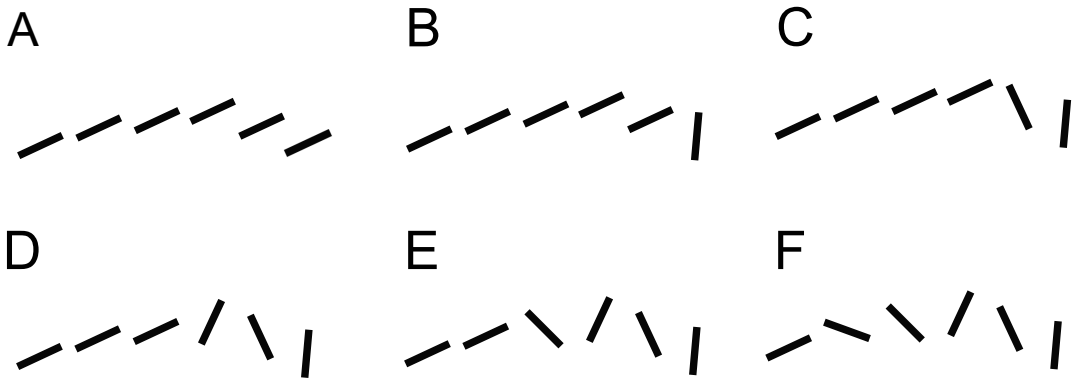} 
\end{center}
\caption{{\bf Under the mixture of Gaussians prior, more homogeneous configurations of orientations are given greater probability.} In the examples shown here, the homogeneity of orientations, and hence the probability under $q(\mathbf{s})$, decreases from A to F. See the main text for details of how the homogeneity preference is implemented in the mixture of Gaussians prior.}
\label{mog_illustrate}
\end{figure}

The model starts with a mixture component that places its probability mass on (or near) a one-dimensional subspace. If $s_i$ denotes the $i^{\rm th}$ stimulus, then this one-dimensional subspace is the line $s_1=s_2=\ldots=s_N$. This mixture component has a covariance matrix of the form given in Equation \ref{Lambda} with a large correlation $\rho$ (we use $\rho = 0.99$). This effectively constrains the component's probability mass to the line $s_1=s_2=\ldots=s_N$, thus implementing a preference for homogeneous stimuli. Using additional mixture components, the model then allocates progressively less probability to higher dimensional subspaces and thus to less homogeneous stimulus configurations. There are, for example, $N$ mixture components that place most of their mass on two-dimensional subspaces with all but one $s_i$ having similar feature values. This is achieved by setting the $i^{\rm th}$ row and the $i^{\rm th}$ column of the correlation matrix (except for the entry on the diagonal which is always equal to 1) to a small value (we use 0, meaning no correlations between $s_i$ and the remaining stimuli) and keeping the remaining correlations high. This process is continued until the $N^{\rm th}$-dimensional space, reducing the probability mass allocated as the dimensionality is increased (Figure \ref{mog_illustrate}). Our mixture of Gaussians prior is similar to the joint feature statistics in the ``dead leaves model'' \cite{pitkow2010}, which is known to be a very good model of natural image patches \cite{zoranweiss2012}. 

The mixture of Gaussians model incorporates only one type of ecologically plausible statistical regularity among the stimuli, namely homogeneity. In a more realistic extension of this model, there would be many other high-probability components in the mixture corresponding to other low-dimensional manifolds or subspaces that have a high probability in the natural visual environment, such as manifolds corresponding to smooth contours or regular shapes for the case of orientation. It seems quite plausible that such additional regularities would increase the mismatch between the subject's prior and the experimenter's prior, which is generally a completely factorized and uniform distribution, and thus amplify the effects of model mismatch considered in this paper. It would be interesting to investigate in more detail the consequences of incorporating such additional regularities in the model, but this is beyond the scope of the current paper. 

Parameter values (i.e., mixture weights of the components) for the mixture of Gaussians model were chosen manually such that (i) components corresponding to less homogeneous configurations were given progressively smaller weights as explained above, and (ii) the resulting model produced significant encoding variability as will be discussed below. To investigate the effects of varying the parameters of the model, we also conducted simulations where the individual weights were exponentiated to a common non-negative exponent $k$ (i.e., $w_j^k$, where $w_j$ denotes the default manually-chosen weight of the $j^{\rm th}$ component) and then renormalized to ensure that the weights sum to 1. Note that $k=1$ corresponds to the default setting of the weights that were manually chosen as described above (Text S1). Smaller $k$ values correspond to more uniform weights and larger $k$ values correspond to sparser weights with most of the total weight concentrated on the most homogeneous component. 

\section*{Results}
To demonstrate that model mismatch alone, without a general resource or capacity limitation, can account for some of the main qualitative characteristics of performance limitations in VSTM tasks, we consider a scenario where the noise distribution $p(\mathbf{x}|\mathbf{s})$ does not depend on the set size (i.e., the marginal variance of the noise distribution along each dimension is constant across different set sizes). This implements the assumption that a general resource or capacity limitation does not play a significant role in the observed performance limitations in VSTM tasks. As discussed in more detail shortly, under a general resource limitation, the variance of the noise distribution would increase with set size. We will also briefly discuss the effects of combining model mismatch with a resource limitation. The complete set of results for the latter scenario is provided in the Supporting Information (Text S5, Figures S5-S8).

\subsection*{Memory precision-set size relationship}
\label{setsize_sec}
A prominent result cited as evidence for a general resource or capacity limitation in VSTM is the monotonic decline in subjects' memory precision with set size in standard VSTM experiments. Most theories explain this decline in memory precision in terms of a limited amount of memory resources. As the number of items increases, the amount of available resources per item decreases, hence the precision with which an individual item can be encoded declines as a power law of the number of items \cite{palmer1990}. Different theories disagree on the nature of resources. One theory claims that the resources are a small number of discrete, fixed-precision memory `slots' \cite{zhangluck2008}, \cite{andersonetal2011}, whereas another theory claims that the resources are continuous \cite{bayshusain2008}, \cite{wilkenma2004}. In one version of the second type of theory, for instance, it is hypothesized that stimuli are encoded by neural populations, and there is a constraint on the amount of neural resources that can be expended to encode a visual display \cite{bayshusain2008}, \cite{mahuang2009}. For example, there might be a limit on the total number of neural spikes \cite{mahuang2009}. In this case, larger set sizes lead to less spikes per item. Under a plausible model of neural variability (i.e. Poisson-like variability), this implies that increases in set size should lead to increases in the variance of $p(\mathbf{x}|\mathbf{s})$ that scale as $N$ \cite{mahuang2009}. This predicted increase in variance with set size is slower than that observed in most VSTM experiments which typically report a power-law relationship between variance and set size with exponents around 1.2-1.5 \cite{bayshusain2008}, \cite{vandenbergetal2012}, \cite{baysetal2011}.  

Here, we show that set size effects can be explained in terms of a mismatch between the experimenter's prior $p(\mathbf{s})$ and the subject's prior $q(\mathbf{s})$ without assuming a general resource limitation (i.e., without assuming an increase in observation noise with set size). This result warrants a reconsideration of the theoretical arguments regarding resource or capacity limitations mentioned in the previous paragraph, because it shows that a general resource limitation is not the only possible explanation of the memory precision--set size relationship. 

\begin{figure}[!ht]
\begin{center}
\includegraphics[scale=1]{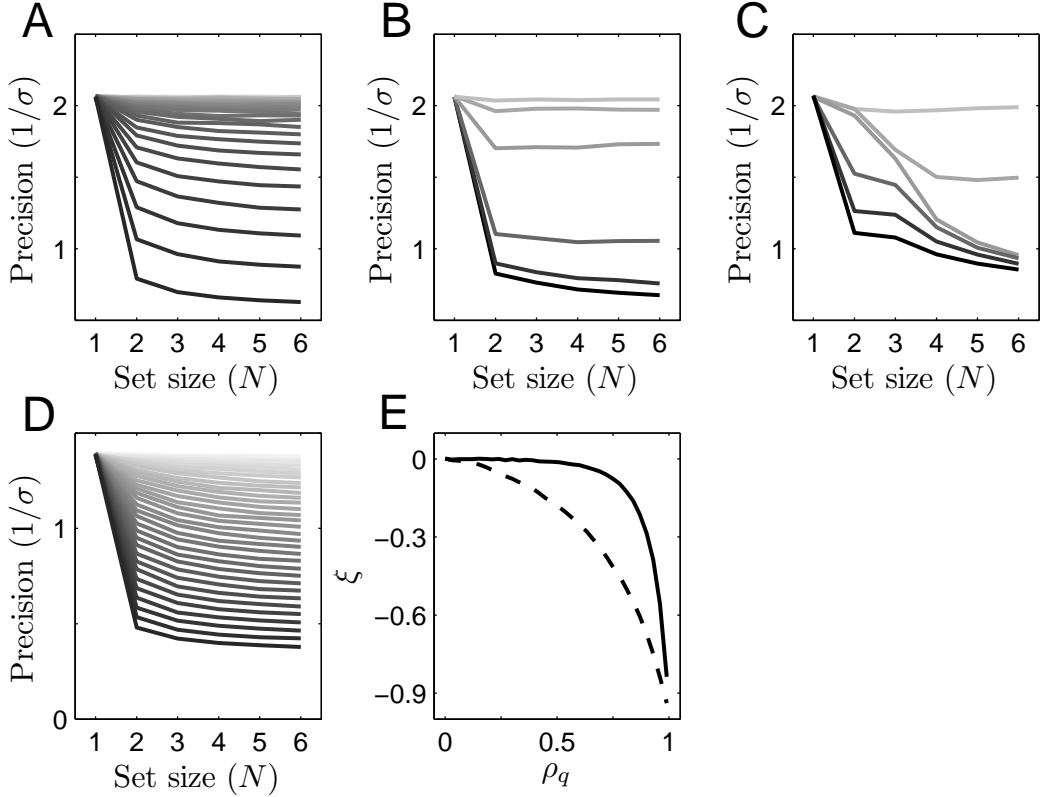} 
\end{center}
\caption{{\bf Recall precision as a function of set size.} (A) Recall precision as a function of set size in the Gaussian model with uniform non-negative correlations. Different curves correspond to different correlation values, denoted $\rho_q$. 34 different $\rho_q$ values are used in this example, from 0 to 0.99 with increments of 0.03. Lighter colors represent lower $\rho_q$ values. Following Bays \& Husain (2008), precision is defined as 1 over the standard deviation of errors. (B) Recall precision as a function of set size in the Gaussian model with random positive correlations. Different curves correspond to different $\gamma$ values. Six different $\gamma$ values are used in this example, $\gamma=0.1,0.15,0.3,0.9,1.8,3.6$. Lighter colors represent higher $\gamma$ values. (C) Recall precision as a function of set size in the mixture of Gaussians model. Different curves correspond to different weight exponents, $k$. Six different $k$ values are used in this example, $k=0,0.2,1,7,13,19$. Lighter colors represent smaller $k$ values. (D) Recall precision as a function of set size in the Gaussian model with uniform correlations across dimensions. In this example, in addition to the difference in correlation, the experimenter's prior and the subject's prior also differ in variances, with the experimenter's prior having a standard deviation in each dimension that is three times the standard deviation of the subject's prior ($\sigma_p = 3$, $\sigma_q = 1$). Different curves correspond to different $\rho_q$ values. As in (A), 34 different $\rho_q$ values are shown, from 0 to 0.99 with increments of 0.03. Lighter colors represent lower $\rho_q$ values. (E) Exponents of power law fits, $1/\sigma \propto N^{\xi}$, to precision-set size curves shown in (A) [solid line] and (D) [dashed line] as a function of $\rho_q$.}
\label{setsize}
\end{figure}

Figure \ref{setsize}A shows the recall precision (inverse standard deviation of the error distribution) as a function of set size for the case of Gaussian $q(\mathbf{s})$ with uniform non-negative correlations and an uncorrelated $p(\mathbf{s})$. Although Figure \ref{setsize}A shows simulation results, in this case, an analytic expression for the precision of the error distribution can be derived in terms of the marginal standard deviations of $p(\mathbf{s})$ and $q(\mathbf{s})$, the correlation coefficient $\rho_q$ of $q(\mathbf{s})$ and the set size $N$ (Text S3). Simulation results are in excellent agreement with the analytic predictions (Figure S1). The full picture that emerges from this analytic expression is surprisingly complicated and the precision set-size relationship is not always monotonically decreasing (Figure S3). However, as a general rule, when the marginal standard deviations of $p(\mathbf{s})$ and $q(\mathbf{s})$ are similar and $\rho_q>0$, the precision is a decreasing function of $N$ for the range of set sizes that are practically relevant (up to about $N \approx 10$) and increasing $\rho_q$ increases this range even further (Figure S3). 

In Figure \ref{setsize}A, simulation results are shown for 34 different correlations $\rho_q$ ranging from 0 to 0.99 with increments of 0.03 and for set sizes from 1 to 6 items. Clearly, the recall precision decreases with set size and this set size effect increases monotonically with $\rho_q$. To quantify the set size effect, we fit power-law functions of the form $cN^{\xi}$ to these data. Smaller exponents indicate more drastic reductions in precision with set size. Empirically, it has been observed that power-law functions with exponents ranging from $-0.60$ to $-0.75$ fit subjects' data well \cite{bayshusain2008}, \cite{vandenbergetal2012}, \cite{baysetal2011}. Figure~\ref{setsize}E (solid line) shows the exponents of the power-law fits as a function of $\rho_q$. $\rho_q = 0.96$ yields an exponent of $\xi=-0.56$ ($R^2=0.89$) and $\rho_q = 0.99$ yields $\xi=-0.84$ ($R^2=0.90$). 

For a given correlation value $\rho_q$, it is possible to produce smaller exponents---that is, more drastic reductions in precision with set size---if a mismatch between the marginal variances of the experimenter's prior and the subject's prior, in addition to the mismatch in the correlation structure, is introduced. Underestimating and overestimating the actual variance have different effects on the precision (Text S3, Figure S2). In general, underestimating the actual variance incurs a heavier cost to precision than overestimating the actual variance (Figure S2). Figure \ref{setsize}D shows precision as a function of set size when the experimenter's prior is assumed to have a marginal standard deviation that is three times the marginal standard deviation of the subject's prior ($\sigma_p = 3$, $\sigma_q = 1$), and Figure~\ref{setsize}E (dashed line) shows the exponents of the corresponding power-law functions.

Next, we consider the subject's prior modeled as a multivariate Gaussian with random positive correlations between stimuli. Figure \ref{setsize}B shows the recall precision as a function of set size in this case. Results are shown for 6 different $\gamma$ values: 0.1, 0.15, 0.3, 0.9, 1.8, 3.6. The set size effect decreases monotonically with $\gamma$. Setting $\gamma=0.15$ yields a power-law exponent of $\xi=-0.67$ ($R^2 = 0.87$). Again, for a given $\gamma$ value, it is possible to produce smaller exponents if $q(\mathbf{s})$ underestimates the marginal variance of $p(\mathbf{s})$.

Finally, Figure \ref{setsize}C shows the results when the subject's prior is modeled using the mixture of Gaussians model. Results are shown for 6 different $k$ values: 0, 0.2, 1, 7, 13, 19. Recall that $k=0$ corresponds to the uniform weighting of all components, $k=1$ corresponds to the default setting of the weights, and larger $k$ values correspond to sparser weights with most of the total weight concentrated on the most homogeneous component. The set size effect increases monotonically with $k$. Different $k$ values also produce qualitatively different shapes for the precision--set size curves. Setting $k=1$ yields a power-law exponent of $\xi=-0.40$ ($R^2 = 0.85$). 

Both for the Gaussian model with uniform non-negative correlations and for the Gaussian model with random positive correlations, especially for large correlation values, a large proportion of the total reduction in recall precision is obtained when the set size is increased from $N=1$ to $N=2$. For the mixture of Gaussians model, on the other hand, substantial reductions in precision can still be observed beyond $N=2$. This is presumably due to the existence of higher-order dependencies between stimuli under the mixture of Gaussians prior, whereas the Gaussian models incorporate only second-order dependencies.

{\color{black}The measure of recall accuracy we use here, i.e. the precision of the error distribution, takes into account both the bias and the variance of the model responses. Biases of the three models for different set sizes are presented separately in the Supporting Information (Text S4, Figure S4).}

Combining model mismatch with a resource limitation (i.e., with a noise distribution whose marginal variance scales as $N$) produces a steeper decline in recall precision than that obtained by either model mismatch or the resource limitation alone (Text S5, Figure S5). For example, in the Gaussian model with uniform correlations, without any model mismatch (i.e., $\rho_p=\rho_q=0$), recall precision decreases as a power-law function of $N$ with an exponent of $-0.44$ (the slightly greater than $-0.5$ exponent is due to the use of a relatively informative prior in our simulations). Model mismatch decreases this exponent further, with larger mismatches leading to smaller exponents. The effects of model mismatch and the resource limitation do not combine additively. For example, without a resource limitation, a correlation coefficient of $\rho_q=0.81$ by itself produces $\xi = -0.12$ ($R^2=0.92$), but combined with the resource limitation, the same amount of model mismatch leads to $\xi = -0.61$ ($R^2=0.99$), a steeper decline in precision than would be predicted if model mismatch and the resource limitation combined additively. 

This result suggests a possible explanation of the steeper than predicted declines in precision observed in VSTM studies. Recall that the theoretical arguments discussed at the beginning this subsection predict the noise variance to scale as $N$, which in turn corresponds to a precision (inverse standard deviation) that scales as $N^{-0.5}$ (assuming a relatively non-informative prior). But, most VSTM studies report power-law relationships between recall precision and set size with exponents ranging from $-0.60$ to $-0.75$ \cite{bayshusain2008}, \cite{vandenbergetal2012}, \cite{baysetal2011}. A possible explanation of this discrepancy is that the additional decline in precision, not explained by the increase in noise variance, might be caused by model mismatch.

\subsubsection*{Correlated $p(\mathbf{s})$}
We next investigate the consequences of introducing dependencies between stimuli in the experimenter's prior $p(\mathbf{s})$. We show that this can lead to an ``inverse set size effect'' in which recall precision increases with set size.

In these simulations, the experimenter's prior $p(\mathbf{s})$ is assumed to be a multivariate Gaussian with uniform non-negative correlations $\rho_p$ between each pair of its dimensions. Setting $\rho_p=0$ results in the case of uncorrelated $p(\mathbf{s})$ discussed previously. 

\begin{figure}[!ht]
\centering
\includegraphics[scale=1]{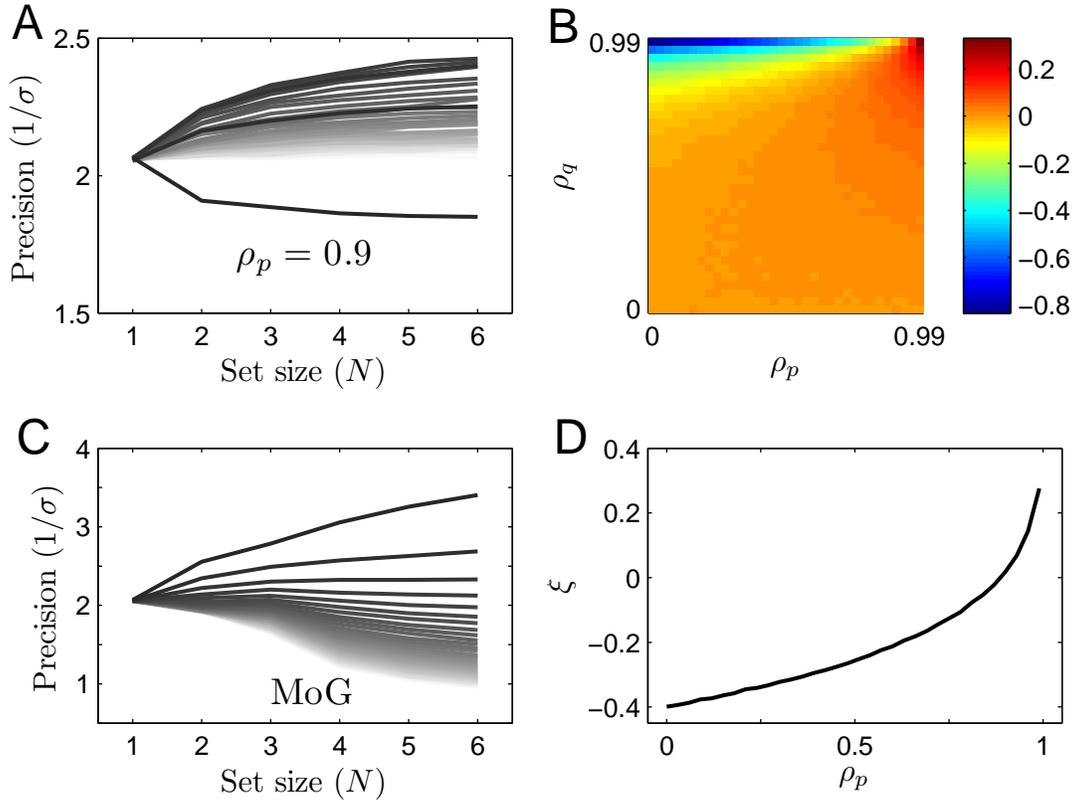} 
\caption{{\bf Recall precision as a function of set size (correlated $p(\mathbf{s})$).} (A) Recall precision as a function of set size in the Gaussian model with uniform non-negative correlations. In these simulations, $p(\mathbf{s})$ as well as $q(\mathbf{s})$ is assumed to be a multivariate Gaussian with uniform correlations. Correlations $\rho_p$ and $\rho_q$ denote the uniform correlations between pairs of dimensions in $p(\mathbf{s})$ and $q(\mathbf{s})$, respectively. $\rho_p$ was set to $0.9$. Precision-set size curves are shown for 34 different values of $\rho_q$ from $0$ to $0.99$ in increments of 0.03. Lighter colors represent lower $\rho_q$ values. Note that when $\rho_q$ exceeds $\rho_p$, the inverse set size effect starts to decrease and ultimately reverses to a set size effect (a decline in precision with set size). (B) Exponents $\xi$ of power-law fits to precision-set size curves for each pair of ($\rho_p$, $\rho_q$) values. Positive values indicate an increase in precision with set size, whereas negative values indicate a decrease in precision with set size. (C) Results for the mixture of Gaussians model. $p(\mathbf{s})$ is a multivariate Gaussian with uniform correlations across dimensions, and $q(\mathbf{s})$ is a mixture of Gaussians with weight exponent set to $k=1$. $\rho_p$ is varied from $0$ to $0.99$ in increments of 0.03. Each curve represents the precision-set size curve for a different $\rho_p$ value. Lighter colors represent lower $\rho_p$ values. (D) Exponents of power law fits, $\xi$, to precision--set size curves in the mixture of Gaussians model shown in (C) as a function of $\rho_p$.}
\label{correlated_pq_fig}
\end{figure}

We first consider the case in which the subject's prior $q(\mathbf{s})$ is a multivariate Gaussian with uniform non-negative correlations $\rho_q$. Again, in this case, an analytic expression for the precision of the error distribution can be derived in terms of the marginal standard deviations of $p(\mathbf{s})$ and $q(\mathbf{s})$, the correlation coefficients of $q(\mathbf{s})$ and $p(\mathbf{s})$ and the set size $N$ (Text S3). Our main finding is that if both $p(\mathbf{s})$ and $q(\mathbf{s})$ display correlations, an inverse set size effect---an increase in recall precision with set size---can be observed. This can be seen in Figure~\ref{correlated_pq_fig}A which shows the simulation results for $\rho_p=0.9$. This figure also shows that when $\rho_q$ significantly overestimates the actual correlation coefficient $\rho_p$, a decline in recall precision with set size (that is, a standard set size effect) is observed even though both $\rho_p$ and $\rho_q$ are positive (e.g., see the darkest line in Figure~\ref{correlated_pq_fig}A). 

Furthermore, the inverse set size effect becomes more pronounced with larger correlations in the experimenter's prior. This is illustrated in Figure~\ref{correlated_pq_fig}B which shows the exponents of power-law fits to precision-set size curves for each pair of ($\rho_p$, $\rho_q$) values tested. Positive exponents indicate an increase in precision with set size, whereas negative exponents indicate a decrease in precision with set size. Overall, the predicted decreases in precision with set size are larger in magnitude than the predicted increases in precision with set size (note the maximum and minimum values of the color bar in Figure~\ref{correlated_pq_fig}B). Results for the case of Gaussian $q(\mathbf{s})$ with random positive correlations were qualitatively similar.

The increase in recall precision with set size can be understood intuitively as follows. When there are dependencies between different stimuli in $p(\mathbf{s})$, it becomes possible to gain information about an individual stimulus, $s_t$, from the remaining stimuli, and the amount of information gained from the remaining stimuli increases as the number of those stimuli increases. This can also be thought of as a ``context effect''---it becomes easier to recall or recognize a stimulus within the context of other stimuli correlated with it, compared to recalling or recognizing it in isolation \cite{bar2004}.

The implication of the predicted increase in recall precision with set size when both $\rho_p$ and $\rho_q$ are positive is that if the experimenter uses a highly correlated prior (e.g., uses highly homogeneous displays in each trial), one would expect to see an increase in subjects' recall precision with increasing set size. However, one has to be careful in translating the results of the simulations directly into experimental predictions. Several caveats apply. First, the magnitude of the increases in recall precision with set size is, in general, smaller than the magnitude of the decreases. Therefore, it may be difficult to design experiments with sufficient power to detect increases in recall precision. Second, the magnitude of the increases in precision depends on the degree of match (or mismatch) between $p(\mathbf{s})$ and $q(\mathbf{s})$. For models more complex than the ones considered in this paper, it is possible for both $p(\mathbf{s})$ and $q(\mathbf{s})$ to display strong dependencies and yet be sufficiently different from each other for the predicted increases in precision to be small or even non-existent (similar to the case represented by the darkest line in Figure~\ref{correlated_pq_fig}A).

Figure \ref{correlated_pq_fig}C shows the results for the mixture of Gaussians prior favoring homogeneity. In this case, the experimenter's prior $p(\mathbf{s})$ is assumed to be a multivariate Gaussian with uniform correlations $\rho_p$ across dimensions, and the subject's prior $q(\mathbf{s})$ is assumed to be a mixture of Gaussians with $k=1$. We varied $\rho_p$ from $0$ to $0.99$ in increments of 0.03. Each curve in Figure~\ref{correlated_pq_fig}C represents the precision-set size curve for a different $\rho_p$ value, with lighter colors representing lower $\rho_p$ values. Again, as in the Gaussian models, for sufficiently high $\rho_p$, an increase in recall precision with set size is predicted, although, for a given $\rho_p$ value, the magnitude of the predicted increase in recall precision is, in general, smaller than in the Gaussian models. Figure~\ref{correlated_pq_fig}D shows the exponents of power law fits to the precision--set size curves in the mixture of Gaussians model as a function of $\rho_p$.

Inverse set-size effects are eliminated when a resource limitation (i.e., a noise distribution whose marginal variance scales as $N$) is introduced (Text S5, Figure S5). This is because the dependencies in the experimenter's prior are no longer strong enough to overcome the effect of increasing noise with set size, although these dependencies predictably lead to shallower precision-set size curves (Figure S5).

To summarize our results on the precision--set size relationship: first, our results show that model mismatch can qualitatively explain the observed decline in memory precision with increasing set size without assuming a general resource limitation. In general, introducing more mismatch between the subject's prior and the experimenter's prior leads to steeper declines in precision with set size (although underestimating vs. overestimating the actual parameters might have asymmetric effects on recall precision). For instance, we found that introducing a mismatch in the marginal variances of these two priors, in addition to a mismatch between their correlation structures, can amplify the set size effect. Second, the precision-set size relationship observed in VSTM tasks might crucially depend on the particular choice for the stimulus distribution $p(\mathbf{s})$ used in an experiment. Our results show that if stimuli are independent under $p(\mathbf{s})$, then it is possible to get a set size effect (Figure~\ref{setsize}A). However, if stimuli are dependent, it is possible to get an inverse set size effect (Figure~\ref{correlated_pq_fig}A). We only considered a multivariate Gaussian form for $p(\mathbf{s})$, which can only model second-order dependencies between stimuli. We expect that additional higher-order dependencies in $p(\mathbf{s})$ would amplify the predicted inverse set size effect.

\subsection*{Variability in {\color{black}memory} precision}
\label{variability_sec}

Recent experimental evidence suggests that {\color{black}memory precision for individual items} in VSTM tasks varies both across trials and across items in a single trial \cite{vandenbergetal2012}, \cite{fougnieetal2012}. Models that incorporate a mixture of multiple precision components, called scale mixtures, generally fit the distribution of recall errors in such tasks significantly better than models with a single precision component such as a single Gaussian distribution. As in set size effects, this result has been interpreted in terms of the properties of the noise distribution $p(\mathbf{x}|\mathbf{s})$, that is, in terms of variability in the noise distribution, such as {random, unstructured} variability in the marginal variances along different dimensions of the noise distribution as well as {random} variability in the variances across trials. The potential source of this {variability in memory precision} is unclear. Instability of neural representations (drift or diffusion of neural responses) due to neural noise \cite{fougnieetal2012} or fluctuations in attention across trials and across items in a single trial \cite{vandenbergetal2012} have been speculated as possible mechanisms that might explain {variability in precision}. Here, we show that these complicated mechanisms might be unnecessary for explaining {variability in precision} in VSTM. Properties of the subject's prior, and interactions between the experimenter's prior and the subject's prior, can naturally account for {variability in memory precision} both across trials and across items within a single trial without hypothesizing any {random} variability in the variance of the noise distribution $p(\mathbf{x}|\mathbf{s})$ itself.

Intuitively, {variability in memory precision} arises according to the perspective proposed here, because when stimuli $\mathbf{s}$ are drawn randomly from the experimenter's prior $p(\mathbf{s})$, in some trials, just by chance, $\mathbf{s}$ will fall near a region where the subject's prior $q(\mathbf{s})$ has high precision. In those trials, the subject will {represent} the stimuli with high precision, {because the posterior distribution will have high precision}. In other trials, however, $\mathbf{s}$ will fall in a region where the subject's prior $q(\mathbf{s})$ has low precision, in which case the subject {represents} the stimuli with low precision, {because the posterior distribution has low precision}. Similarly, in different trials, $\mathbf{s}$ might fall in regions that have high precision along some dimensions according to $q(\mathbf{s})$, but low precision in other dimensions, producing within-trial variability in {memory precision}. 

There is no straightforward relationship between the degree of mismatch between $p(\mathbf{s})$ and $q(\mathbf{s})$ and the amount of variability in {memory precision}. {Variability in precision} might arise even when the experimenter's prior exactly matches the subject's prior. Imagine, for example, that the subject's prior is a mixture of Gaussians with two components, one highly correlated, and the other uncorrelated. If the experimenter's prior exactly matches the subject's prior, in some trials, stimuli will fall in a region where the correlated component has high probability. In those trials, {posterior precision} will be higher. In other trials, however, stimuli will fall in a region where the uncorrelated component has high probability, leading to a lower {posterior precision}. Overall, there will be variability in precision despite the perfect match between the experimenter's and the subject's priors. Conversely, variability in precision may be low, or even non-existent, even when there is a mismatch between the experimenter's and subject's priors. Again, imagine that the subject's prior is a mixture of a highly correlated Gaussian and an uncorrelated Gaussian. If samples from the experimenter's prior consistently come from a region where the correlated component in the subject's prior has low probability, the stimuli will be attributed to the uncorrelated component and the variability in precision will be low.

For a given set size, variability in {memory precision} does not arise in the Gaussian $q(\mathbf{s})$ with uniform non-negative correlations. It is easy to see why this is the case. The same prior with a homogeneous covariance matrix (Equation \ref{Lambda}) is used in each trial, resulting in a posterior distribution with the same homogeneous covariance matrix in each trial, meaning that individual stimulus features have identical marginal variances in each trial. Thus, we only consider the remaining two models here.

To show that the other two models generate variability in {memory precision}, we simulate a large number of trials and fit the error distributions produced by each of these models with both a single precision distribution (a single Gaussian distribution) and an infinite mixture of Gaussians with different precisions or, more technically, a scale mixture of Gaussians with a Gamma distribution over the precisions (where ``precision'' is now used in the sense of inverse variance). The latter distribution is also known as the $t$-distribution, and it has heavier tails and a more prominent peak than a Gaussian with the same mean and variance. As mentioned previously, empirical error distributions from VSTM experiments are generally significantly better fit by a scale mixture distribution than by a single precision distribution, indicating the presence of variability in {memory precision}. To compare the fits of these two distributions to the error distributions produced by our models, we employ the Bayesian information criterion (BIC) \cite{schwartz1978}. The single precision model has two free parameters (mean and variance parameters) and the $t$-distribution has three free parameters (location, scale and degrees-of-freedom parameters). For the $t$-distribution, it is possible to infer the implied Gamma distribution over the precisions from the estimated parameters of the $t$-distribution (Text S2). 

\begin{figure}[!ht]
\centering
\includegraphics[scale=1]{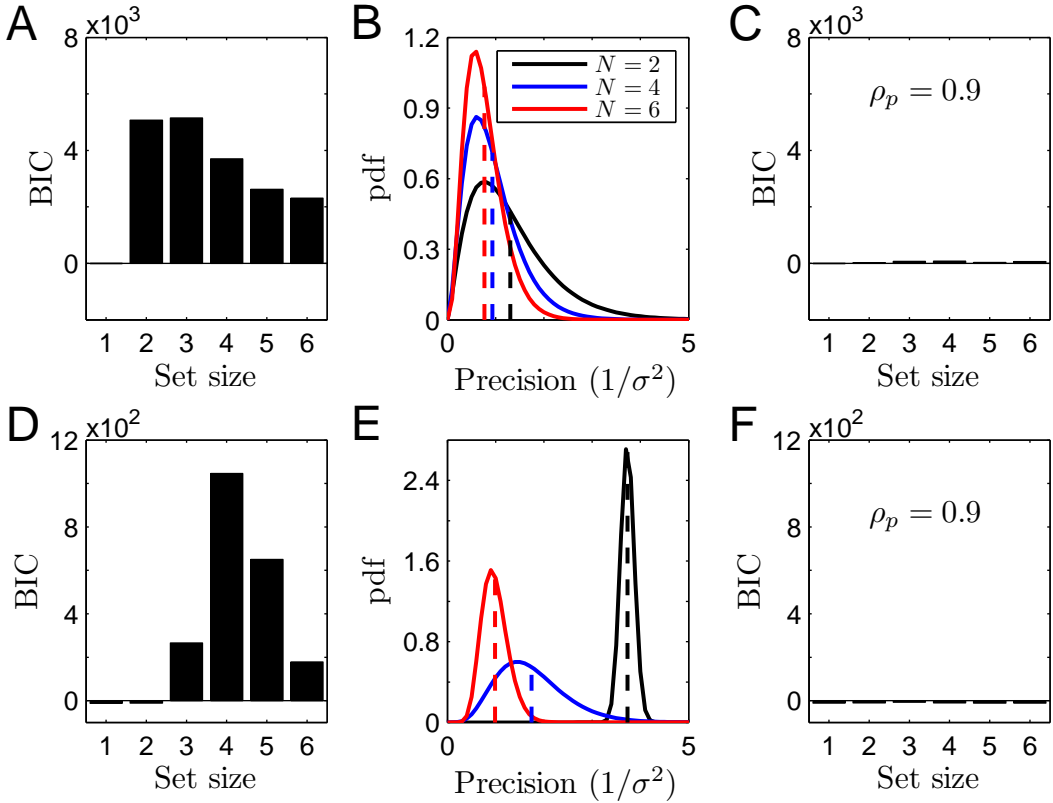} 
\caption{{\bf Variability in memory precision.} (A-B) Results for the case of uncorrelated Gaussian $p(\mathbf{s})$ and Gaussian $q(\mathbf{s})$ with random positive correlations ($\gamma = 0.15$):  (A) BIC score of the Gaussian fit relative to the BIC score of the $t$-distribution fit for each set size. Lower scores indicate better fits. Positive values indicate better fits for the $t$-distribution than for the Gaussian distribution. (B) For three different set sizes, the estimated Gamma distributions over precision for the best fit $t$-distributions. The vertical lines show the means of the Gamma distributions. (C) Similar to (A), but here the experimenter's prior contains correlations ($\rho_p=0.9$). (D-E) Similar to (A-B) except the results are for the case of uncorrelated Gaussian $p(\mathbf{s})$ and the mixture of Gaussians $q(\mathbf{s})$ that favors homogeneity ($k=1$). (F) Similar to (D), but the experimenter's prior contains correlations ($\rho_p=0.9$).}
\label{variability}
\end{figure}

As before, the noise distribution $p(\mathbf{x}|\mathbf{s})$ is assumed to be a multivariate Gaussian with an isotropic, diagonal covariance matrix (variance along each dimension is equal to 0.25 as in the previous simulations). Importantly, the noise distribution itself does not contain any variability (i.e., the marginal variances of the noise distribution do not vary across items or trials). For the experimenter's prior, $p(\mathbf{s})$, we first use the uncorrelated Gaussian distribution (with $\rho_p=0$ and $\sigma_p^2=4$). Figures~\ref{variability}A and~D show the BIC scores of the single precision Gaussian distribution relative to the BIC score of the $t$-distribution for different set sizes. Positive values indicate better fits for the $t$-distribution. For the Gaussian $q(\mathbf{s})$ with random positive correlations (Figure \ref{variability}A), all set sizes except for the set size of 1 indicate better fits for the $t$-distribution, suggesting the presence of multiple precision components (the concentration parameter $\gamma$ was set to 0.15 in these simulations). For the mixture of Gaussians prior favoring homogeneity with $k=1$ (Figure \ref{variability}D), all set sizes other than 1 and 2 indicated better fits for the $t$-distribution, again suggesting the presence of multiple precision components (all positive BIC values were larger than 150 indicating very strong evidence in favor of the $t$-distribution). Figures~\ref{variability}B and E show the inferred Gamma distributions over precision for the best fit $t$-distributions. Predictably, larger set sizes lead to smaller mean precisions. 

To investigate the effects of correlations in the experimenter's prior, we next use a multivariate Gaussian $p(\mathbf{s})$ with a uniform correlation of $\rho_p=0.9$ between each pair of dimensions. Introducing correlations in $p(\mathbf{s})$ significantly reduced variability in {memory precision} for the case of Gaussian $q(\mathbf{s})$ with random positive correlations (Figure \ref{variability}C) and completely eliminated variability in {memory precision} for the mixture of Gaussians model (Figure \ref{variability}F), although the results for the mixture of Gaussians model were sensitive to the variance of the noise distribution (Text S5).  

Combining model mismatch with a resource limitation yields qualitatively very similar results in the Gaussian model with random correlations, but may yield different results in the mixture of Gaussians model depending on the noise level (Text S5, Figure S6).

The results presented in this subsection have two important implications. First, they show that it is not necessary to hypothesize variability in the noise distribution to explain variability in {memory precision}. Variability in {memory precision} can arise naturally if the subject's prior is sufficiently structured. Consider, for instance, the mixture of Gaussians prior favoring homogeneity. {Variability in precision} arises in this model, because in some trials, the presented stimuli are more homogeneous and hence, under the subject's prior, fall near more correlated components, which leads to a {posterior with higher precision}. Within--trial variability in precision can be explained similarly. A stimulus-dependent noise distribution due to adaptation to natural stimulus statistics would increase the {variability in precision} and might be necessary to account for the experimentally observed {variability in precision} even for $N=1$ \cite{fougnieetal2012} (see below). 

Our results thus suggest that the variability in {memory precision} observed in VSTM tasks might be due, at least in part, to variability in the homogeneity level of the display, or in general variability in the ``naturalness'' of the display under the subject's prior---by chance, the display, or parts of the display, will look more or less natural in different trials. This, in turn, results in the {representation} of items with more or less precision. There is already some experimental evidence for this account of variability in {memory precision} in VSTM tasks \cite{orhanjacobs2013}, \cite{bradyalvarez2012}, \cite{bradytenenbaum2013}. However, it remains to be seen whether this account can explain the full extent of variability in {memory precision} observed in VSTM tasks or an additional stochastic source of variability is needed to account for the experimental results. 

Second, as with the precision-set size relationship discussed in the previous subsection, the results reported in this subsection emphasize the potential importance of the stimulus distribution $p(\mathbf{s})$ for determining patterns in subjects' responses. If stimuli are independent, subjects may show significant variability in {memory precision}. If stimuli are dependent, on the other hand, variability in {memory precision} may be significantly reduced or even eliminated (Figure~\ref{variability}).

\subsection*{Different set-size dependencies for initial encoding rate and asymptotic precision}
\label{encodingrate_sec}

Bays and colleagues \cite{baysetal2011} recently investigated the time course of recall precision in VSTM by varying the presentation time of the visual display in a standard VSTM recall experiment. In each trial in Experiment~1 of \cite{baysetal2011}, a visual display consisting of a number of colored bars was presented, followed by a mask and then a delay period. One of the bars was then cued, and subjects recalled the orientation of the cued bar. Importantly, presentation time of the visual display was varied across trials. Bays et al. \cite{baysetal2011} found that recall precision increased with presentation duration. The relationship between recall precision and presentation time was well-fit by a negative exponential function such that an initial rapid increase in precision is followed by saturation at longer presentation times \cite{baysetal2011}:
\begin{equation}
P(t) = P_{max}(1 - \exp (-t / \tau))
\label{precision_time}
\end{equation}  
where $P(t)$ is the recall precision (inverse standard deviation of the error distribution) for a presentation duration of $t$ and $P_{max}$ denotes the maximum precision. When data from VSTM recall experiments that systematically vary presentation time for different set sizes are fit with a function of the form given in Equation \ref{precision_time}, a power-law relationship is observed between set size $N$ and $P_{max}$, with an exponent around $-0.60$ \cite{baysetal2011}. This is the familiar set size effect discussed previously. However, initial encoding rate, $P_{max}/\tau$, or the slope of the initial part of the negative exponential function in Equation \ref{precision_time}, when fit to behavioral data, displays a $1/N$ relationship to set size \cite{baysetal2011}, suggesting that $\tau$ should be roughly proportional to $N^{0.4}$. This implies that increasing the set size affects the initial rate of information accrual into VSTM more than it affects the asymptotic precision. The stronger dependence of initial encoding rate, compared to asymptotic precision, on set size led Bays et al. to suggest that there is an additional cost for initially encoding information into VSTM above and beyond the cost of simply storing information in VSTM.

Here, we show that it is unnecessary to assume a set-size dependent noise distribution to explain the results from \cite{baysetal2011}. Specifically, we show that the data can be qualitatively explained if it is assumed that the precision of the noise distribution $p(\mathbf{x}|\mathbf{s})$ changes with presentation time $t$ according to Equation \ref{precision_time}, but is independent of set size (i.e., $P_{max}$ and $\tau$ do not depend on the set size). Given this assumption, different set-size dependencies for asymptotic precision and initial encoding rate are due to an interaction between two factors, the mismatch between the experimenter's and the subject's priors and the time-varying, but set size--independent, noise distribution. This result has important implications for the interpretation of the experimental data. In particular, it suggests that it may not be necessary to assume an extra cost for encoding information into VSTM in addition to the cost of simply storing information in VSTM as hypothesized by Bays et al. The apparent extra cost may rather be due to the mismatch between the experimenter's and the subject's priors. 

\begin{figure}[!ht]
\centering
\includegraphics[scale=1]{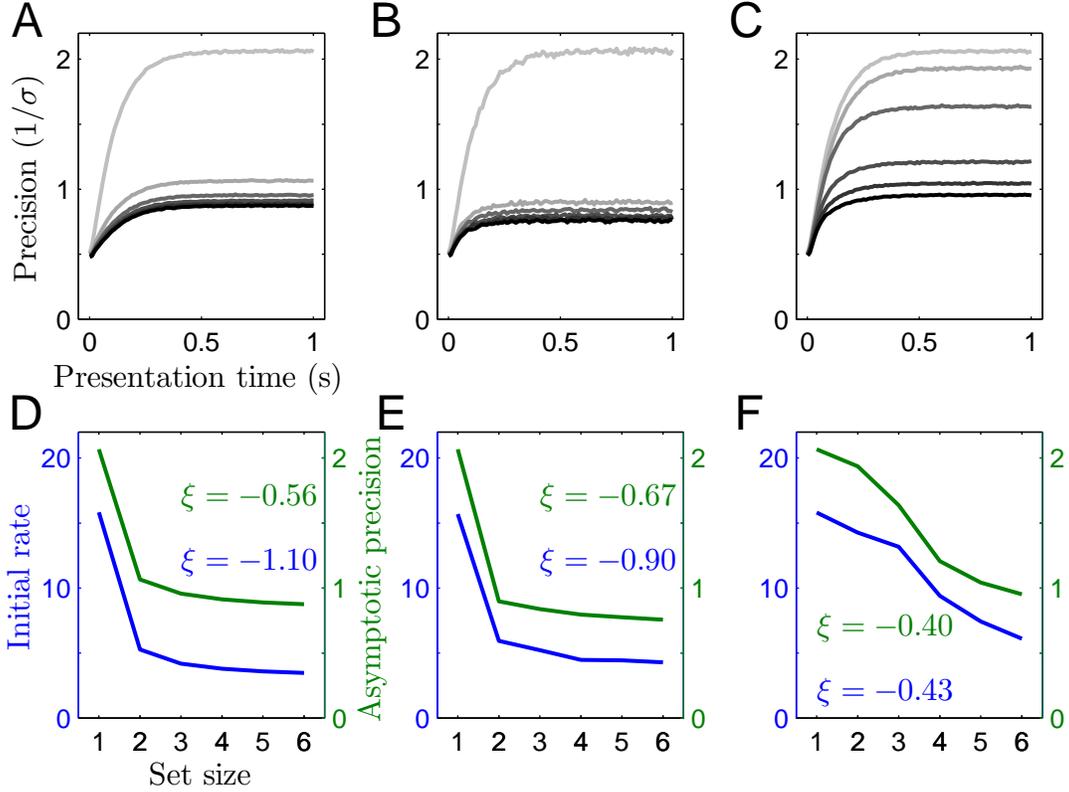} 
\caption{{\bf Recall precision as a function of presentation time.} (A-C) Recall precision as a function of presentation time for different set sizes. Darker colors indicate larger set sizes with set sizes ranging from 1 to 6. Simulations were conducted with the experimenter's prior $p(\mathbf{s})$ set to an uncorrelated Gaussian distribution ($\rho_p=0$). Results are shown for the cases of Gaussian $q(\mathbf{s})$ with uniform correlations [$\rho_q=0.96$] (A), Gaussian $q(\mathbf{s})$ with random positive correlations [$\gamma=0.15$] (B), and the mixture of Gaussians prior favoring homogeneity [$k=1$] (C). (D-F) Initial encoding rate (blue) and the asymptotic precision (green) as a function of set size. The exponents $\xi$ of power-law fits to each curve are also provided. Results are again shown for the case of Gaussian $q(\mathbf{s})$ with uniform correlations (D), Gaussian $q(\mathbf{s})$ with random positive correlations (E), and the mixture of Gaussians prior favoring homogeneity (F). Due to computational demands, $10^4$ trials per presentation time per set size were simulated for the Gaussian model with random positive correlations. For the other two models, $10^5$ trials per presentation time per set size were simulated. The simulated presentation times varied from 5 ms to 1005 ms in steps of 10 ms. The noise distribution has an asymptotic precision of 2 and $\tau=0.1$ for all models.}
\label{encodingrate_fig}
\end{figure}

We start by considering the situation in which the experimenter's prior is an uncorrelated Gaussian distribution ($\rho_p=0$). Figure~\ref{encodingrate_fig}A-C shows recall precision as a function of presentation time for different set sizes. Results are shown for the cases in which the subject's prior $q(\mathbf{s})$ is a Gaussian distribution with uniform non-negative correlations (Figure~\ref{encodingrate_fig}A), a Gaussian distribution with random positive correlations (Figure~\ref{encodingrate_fig}B), and a mixture of Gaussians distribution favoring homogeneity (Figure~\ref{encodingrate_fig}C). We fit the curves shown in Figure~\ref{encodingrate_fig}A-C with negative exponential functions of the form given in Equation \ref{precision_time} with the addition of a constant vertical offset $b$ (this is necessary because, with the relatively informative priors used in our examples, the encoding precision does not drop to 0 for time 0; instead it drops to some positive value determined by the marginal prior variance). From these fits, the initial encoding rate is calculated as $P_{max}/\tau$ and the asymptotic precision is calculated as $P_{max}+b$. Figure~\ref{encodingrate_fig}D-F plots the estimated initial encoding rates (blue) and the asymptotic precisions (green) as a function of set size for the three models. Also shown are the exponents $\xi$ of power-law fits to each curve. 

\begin{figure}[!ht]
\centering
\includegraphics[scale=1,trim = 0mm 0mm 0mm 0.5mm, clip]{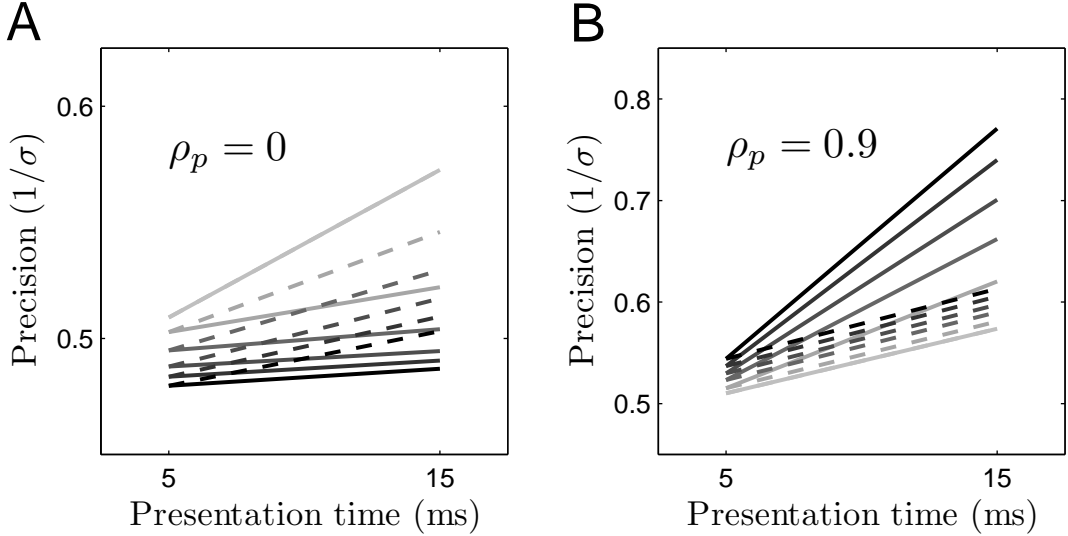} 
\caption{{\bf Blown-up views of precision in the first two time steps of the simulation (corresponding to presentation times of 5 ms and 15 ms respectively) for the Gaussian $q(\mathbf{s})$ with uniform correlations ($\rho_q=0.96$).} Different colors correspond to different set sizes with darker colors representing larger set sizes. Solid lines show the actual simulation results and dashed lines show the precisions that would be predicted if the initial encoding rate had the same dependence on set size as the asymptotic precision (taking the precision for $N=1$ as the baseline). (A) shows the results for the case where the experimenter's prior is an uncorrelated Gaussian ($\rho_p=0$) as in Figure~\ref{encodingrate_fig}A and (B) shows the results for the case where the experimenter's prior is a Gaussian with uniform correlations ($\rho_p=0.9$) as in Figure~\ref{encodingrate_corr_fig}A.}
\label{encodingrate_2_fig}
\end{figure}

Figure~\ref{encodingrate_2_fig}A shows a blown-up view of Figure~\ref{encodingrate_fig}A focusing on the shortest two presentation times (5 ms and 15 ms). Also shown in Figure~\ref{encodingrate_2_fig}A with dashed lines are the precisions that would be predicted if the initial encoding rate had the same dependence on set size as the asymptotic precision (taking the precision for $N=1$ as the baseline). Comparing the slopes of the solid and the dashed lines, it can be seen that the initial encoding rate declines faster than would be predicted if it had the same dependence on set size as the asymptotic precision.  

Importantly, all models produced different set size dependencies for initial encoding rate and asymptotic precision (Figure~\ref{encodingrate_fig}D-F), and these dependencies are qualitatively consistent with the experimental results from \cite{baysetal2011}, that is, the initial encoding rate declines faster than the asymptotic precision as a function of set size. These results were achieved using the assumption that observation noise declines with increasing image presentation times, but does not depend on the set size. The results suggest that there might not be an inherent additional cost to ``writing'' information into VSTM beyond the cost of storing that information in VSTM. Rather, these apparent costs may be a by-product of using experimental stimuli that do not match the subject's prior.

\begin{figure}[!ht]
\centering
\includegraphics[scale=1]{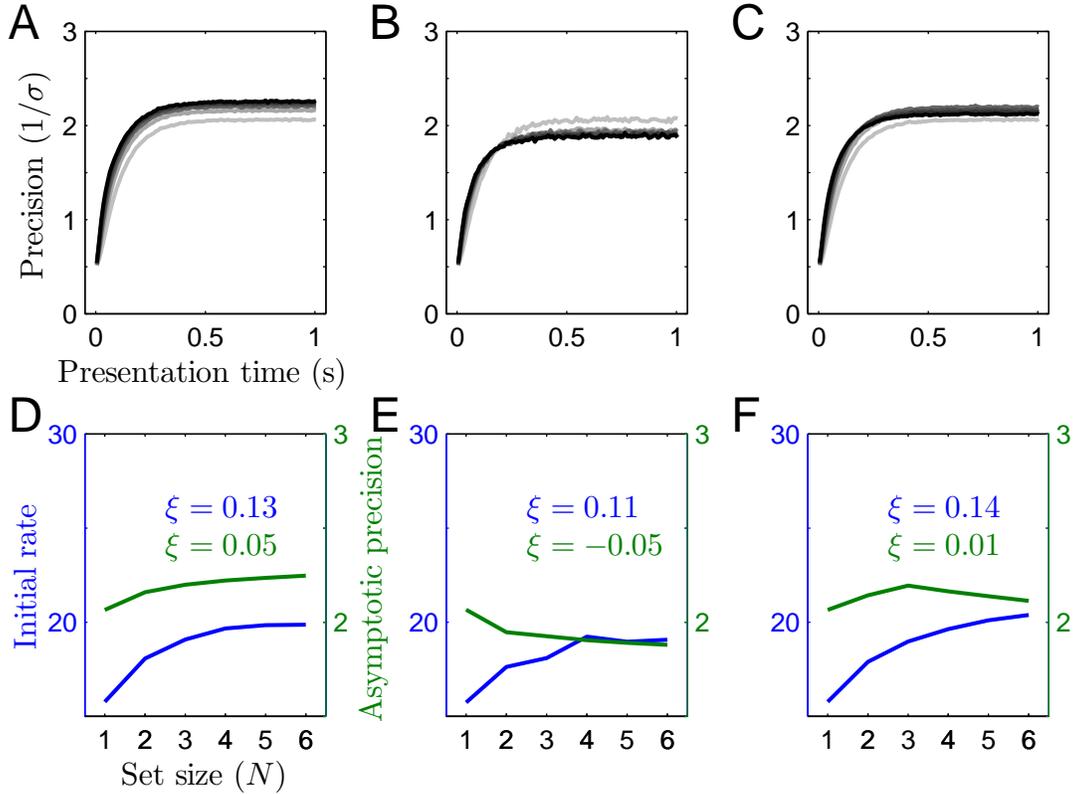} 
\caption{{\bf Recall precision as a function of presentation time.} Results for the case where the experimenter's prior $p(\mathbf{s})$ was a Gaussian distribution with uniform correlations ($\rho_p=0.9$). The formats of the plots are identical to those in Figure~\ref{encodingrate_fig}.}
\label{encodingrate_corr_fig}
\end{figure}

Next, we investigate the effects of correlations in the experimenter's prior $p(\mathbf{s})$ on initial encoding rate and asymptotic precision. This prior is set to a multivariate Gaussian with uniform correlations ($\rho_p = 0.9$). Figure \ref{encodingrate_corr_fig} shows the results of these simulations. As demonstrated previously, the asymptotic precision-set size curves become shallower when $p(\mathbf{s})$ is correlated. For all models, an inverse set size effect is observed for the initial encoding rate, meaning that encoding rate increases with set size. Surprisingly, the power-law exponents for the initial encoding rate are greater than the power-law exponents for the asymptotic precision for all three models (recall that for uncorrelated $p(\mathbf{s})$, the opposite is true---the initial encoding rate declines more steeply with set size than the asymptotic precision). This can also be seen in Figure~\ref{encodingrate_2_fig}B which shows a blown-up view of Figure~\ref{encodingrate_corr_fig}A focusing on the shortest two presentation times. The slopes of the solid lines increase faster than would be predicted if the initial encoding rate and the asymptotic precision had the same dependence on set size. Thus, when a correlated $p(\mathbf{s})$ is used, relative to just maintaining items in VSTM, there is an extra benefit to initially encoding them into VSTM rather than an extra cost.

For the Gaussian model with uniform correlations, an analytic expression for the encoding rate at any time point $t$ can be obtained using the analytic expression for the precision of the error distribution (Text S6). The analytic expression allows us to compute the set-size dependence of the initial encoding rate as a function of the correlation coefficient of the subject's prior both for uncorrelated $p(\mathbf{s})$ and for correlated $p(\mathbf{s})$ (Figure S9). The results reported here are all simulation results. The simulation results are in good agreement with the analytic predictions (Figure S9).

Combining model mismatch with a resource limitation produces qualitatively similar results to those obtained with model mismatch alone for all three models (Text S5, Figures S7-S8).

In summary, our results suggest that the apparent cost to the initial encoding of items may be a by-product of the mismatch between the subject's prior and the experimenter's prior, rather than being an inherent cost. Secondly, our results suggest that the observed declines in asymptotic precision and initial encoding rate with set size may be reduced or even reversed if more ecologically realistic priors (e.g., priors with dependencies among stimuli) are used in VSTM studies. Similarly, our framework predicts that, when more ecologically realistic stimulus distributions are used in VSTM studies, the apparent extra cost to initially encoding items in VSTM compared to just storing them in VSTM may also be either reduced or even reversed. 

\subsection*{The effects of model mismatch become more significant with stimuli exhibiting strong statistical regularities}

\begin{figure}[!ht]
\centering
\includegraphics[scale=1.0,trim = 0mm 0mm 0mm 0.5mm, clip]{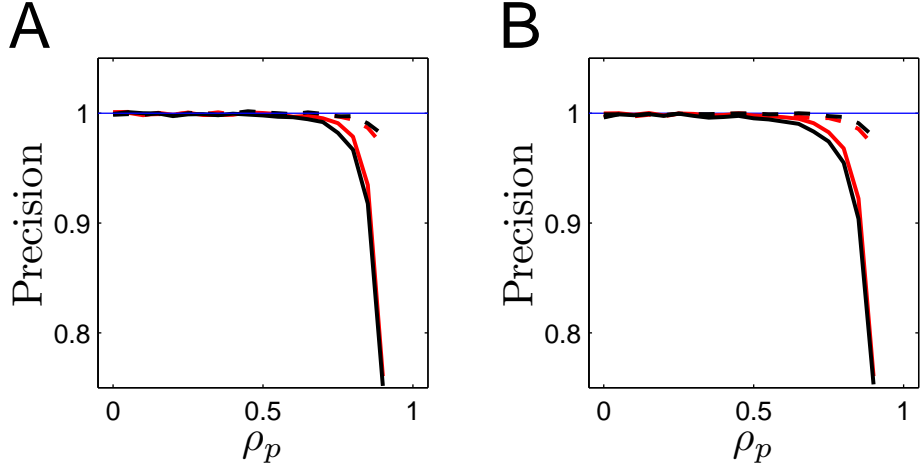} 
\caption{{\bf Precision (as a fraction of optimal precision) as a function of $\rho_p$ in the Gaussian model with uniform correlations.} (A) Dashed lines show the simulation results when the correlation coefficient is underestimated 10\% under the subject's prior. Solid lines show the results when the correlation coefficient is overestimated 10\% under the subject's prior. Red lines show the results when only the correlation coefficient, $\rho_p$, is misestimated. Black lines represent a scenario where there is an additional 10\% underestimation of the variance, $\sigma^2_p$, of the stimulus distribution. (B) is similar to (A), except that the correlation coefficient is under- or over-estimated by a constant value (0.09) under the subject's prior, rather than by a constant proportion of the actual value. Blue line represents the optimal precision.}
\label{strong_corr_demo}
\end{figure}

Do the effects of model mismatch depend on the strength of statistical regularities in the stimulus distribution, $p(\mathbf{s})$? Is model mismatch equally detrimental to performance in domains exhibiting strong statistical regularities as in domains with weak statistical regularities? To address these questions, we performed simulations where $p(\mathbf{s})$ was a Gaussian with uniform correlations $\rho_p$ and the subject's prior under- or over-estimated the actual correlation coefficient either by 10\% (Figure~\ref{strong_corr_demo}A) or by a constant amount (Figure~\ref{strong_corr_demo}B). In both cases, when performance is measured as the recall precision relative to the optimal precision (which would be achieved if the subject's prior used the true parameter values), model mismatch is seen to have a substantially larger impact on performance in domains with stronger statistical regularities (i.e. larger $\rho_p$). This suggests that learning a good model of the environment becomes more important for memory performance in environments with stronger statistical regularities.  

\subsection*{Possible changes in the noise distribution due to adaptation to natural stimulus statistics}
\label{adap_sec}

So far, we have assumed that consequences of adaptation to natural scene statistics are confined to the prior distribution or, in other words, that adaptation to natural scene statistics only changes the prior distribution. However, adaptation can change the noise distribution as well. The essential idea is that if a stimulus is common in the natural visual environment, it should be encoded with better precision under the noise distribution at the expense of encoding less likely stimuli with worse precision. This corresponds to a stimulus-dependent expression for the variance of the noise distribution, $\sigma^2(\mathbf{s})$, with smaller $\sigma^2(\mathbf{s})$ for more likely stimuli $\mathbf{s}$ and larger $\sigma^2(\mathbf{s})$ for less likely stimuli. This is similar to the stimulus-dependent noise distributions found in \cite{girshicketal2011}. This previous work shows that when psychophysical data from simple 2AFC discrimination experiments are fit with models assuming flexible, semi-parametric forms for the noise and prior distributions, the noise variance is found to be smaller in regions where the prior is high. This, in turn, explains the higher sensitivity in the stimulus space around regions where the prior probability is high. Wei and Stocker \cite{weistocker2012} note that a smaller noise variance for more likely stimuli is consistent with the efficient coding hypothesis as it maximizes the information between the actual stimuli and their noisy measurements subject to a resource limitation. Ganguli and Simoncelli \cite{gangulisimoncelli2010} give a neural implementation of this idea (for a one-dimensional stimulus space) by computing the optimal allocation of gains and densities for tuning functions in a neural population encoding a single one-dimensional stimulus. 

Therefore, adaptation to natural scene statistics can manifest its effects both in the prior distribution (by increasing the prior probability of more likely stimuli $\mathbf{s}$) and in the noise distribution (by decreasing the noise variance $\sigma^2(\mathbf{s})$ for more likely stimuli). Stocker and Simoncelli \cite{stockersimoncelli2005} argue that a stimulus-dependent noise variance and its adaptation to stimulus statistics are necessary to account for certain repulsive biases (i.e. biases away from the adapted stimulus) observed in several adaptation studies. Motivated by these studies that suggest the potential importance of the effects of adaptation in the noise distribution, here we investigate the consequences of such adaptation in the noise distribution. 

Specifically, we assume an isotropic diagonal noise distribution $p(\mathbf{x}|\mathbf{s})$ with variance along each dimension given by:
\begin{equation}
\sigma^2(\mathbf{s}) = \frac{f(N)\sigma_0^2}{\delta + q(\mathbf{s})^{1/N}} 
\label{variance}
\end{equation}
where $q(\mathbf{s})$ is the probability density function of the subject's prior, $\delta$ is a positive real number that prevents the denominator from becoming too small and hence determines the largest possible variance, and $f(N)$ is a function of $N$ that determines how the noise variance depends on the set size. The exponent $1/N$ in $q(\mathbf{s})^{1/N}$ is necessary to counteract the automatic normalization that occurs in the density when the dimensionality of $\mathbf{s}$, i.e. the set size $N$, changes. Intuitively, the volume of the effective stimulus space increases as $l^N$ when the stimuli are independent (where $l$ is the length of the effective stimulus range in the one-dimensional case) and the densities become correspondingly smaller. The exponent $1/N$ approximately cancels this dependence on $N$ (the denominator becomes independent of $N$ only when $q(\mathbf{s})$ does not have any dependencies and is only approximately independent of $N$ otherwise, because the volume of the effective stimulus space increases slower than $l^N$ when there are dependencies between dimensions). This is desirable to make $f(N)$ the only set size dependent term in Equation~\ref{variance}. Different choices for $f(N)$ correspond to different assumptions about the relationship between the noise variance and set size. $f(N)=N$ corresponds to a linear increase in the noise variance with set size and implements the assumption of a fixed amount of memory resources. Here, we focus on the case $f(N)=1$ which corresponds to an approximately set-size independent noise variance and implements the assumption that the amount of memory resources increases approximately as $N$. Note that the implicit set-size dependence of $q(\mathbf{s})$ is approximately canceled by the exponent $1/N$ as explained above.

\begin{figure}[!ht]
\centering
\includegraphics[scale=1.0]{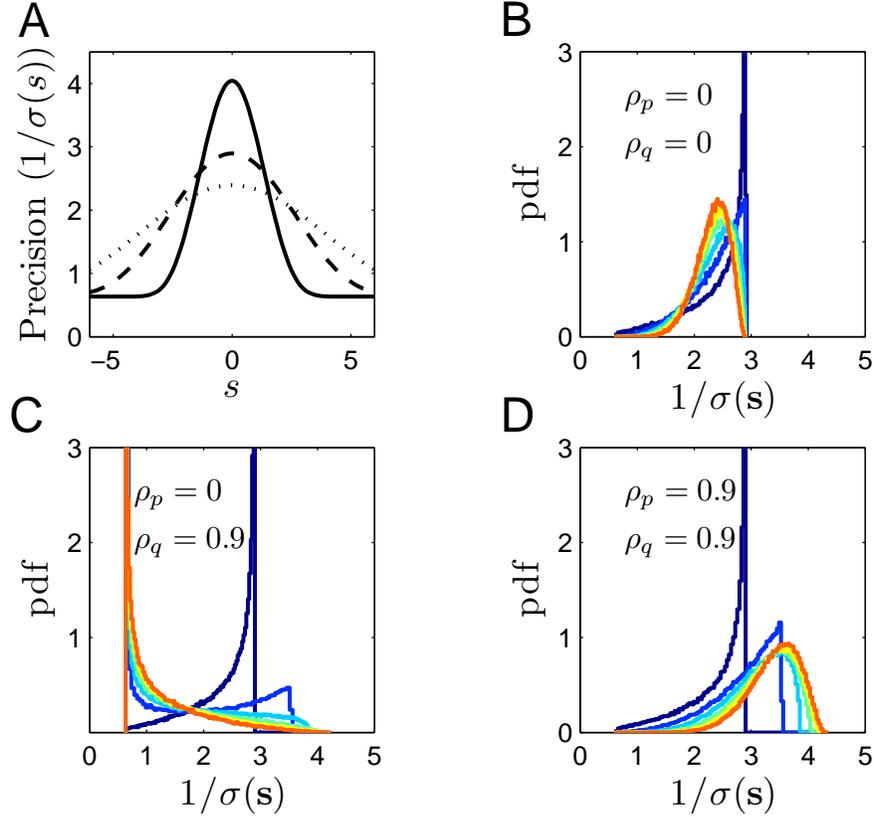} 
\caption{{\bf Stimulus-dependent noise.} (A) A simple demonstration of stimulus-dependent noise in a one-dimensional example. The noise precision is computed from Equation~\ref{variance} where $q(s)$ is taken to be a Gaussian distribution with zero mean and standard deviation equal to either 1 (solid line), 2 (dashed line) or 3 (dotted line). Parameter values are $\sigma_0^2 = 0.025$ and $\delta = 0.01$. (B-D) Empirical distributions of the noise precision over $10^5$ simulated trials for different set sizes from 1 to 6. Hotter colors represent larger set sizes. In (B), both the subject's prior $q(\mathbf{s})$ and the experimenter's prior $p(\mathbf{s})$ are uncorrelated Gaussians with standard deviation 2 along each dimension. In (C), $p(\mathbf{s})$ is an uncorrelated Gaussian with standard deviation 2 along each dimension, but $q(\mathbf{s})$ is a Gaussian with uniform correlations across its dimensions ($\rho_q=0.9$ and $\sigma_q=2$). In (D), both $p(\mathbf{s})$ and $q(\mathbf{s})$ are multivariate Gaussians with uniform correlations ($\rho_p=\rho_q=0.9$ and $\sigma_p=\sigma_q=2$).}
\label{demo_stim_dependent_noise_fig}
\end{figure}

It is easy to see that Equation~\ref{variance} works as desired: for stimuli that have high probability under the subject's prior, the denominator is large, hence the noise variance, $\sigma^2(\mathbf{s})$, is small; conversely, for stimuli that have low probability under the subject's prior, the denominator is small, leading to a large noise variance (also see Figure~\ref{demo_stim_dependent_noise_fig}A). Although Equation~\ref{variance} qualitatively works as desired, the specific form chosen in that equation is \textit{ad hoc}. Ideally, one might want to find out the optimal expression for the noise variance using, for example, population coding ideas as in \cite{gangulisimoncelli2010}. Similarly, Equation~\ref{variance} entails a diagonal noise distribution, but one can easily imagine generalizations with correlated noise. These are directions for future work.

If the experimenter's prior is $p(\mathbf{s})$, ideally the subject should adapt their noise distribution (Equation~\ref{variance}) to $p(\mathbf{s})$: they should encode stimuli that are more likely under $p(\mathbf{s})$ with better precision. If, for example, $p(\mathbf{s})$ is uniform over a given range, $\sigma^2(\mathbf{s})$ should also be uniform. However, following the reasoning laid out in the Methods section, we assume that subjects cannot always achieve this optimal allocation of resources. Instead their noise distribution might reflect adaptation to a different distribution $q(\mathbf{s})$ that may, for example, correspond to the statistics of $\mathbf{s}$ in the natural environment. This leads to suboptimal allocation of resources (within the context of the experiment), because intuitively, the subject is ``wasting'' their resources by encoding $\mathbf{s}$ that have high probability under $q(\mathbf{s})$ (e.g. homogeneous configurations or configurations that correspond to smooth contours) with high precision at the expense of the remaining $\mathbf{s}$ and yet these configurations might have very small probability under $p(\mathbf{s})$. We refer to this situation as suboptimal encoding. Note that this type of suboptimal encoding can occur even if the resources increase with $N$. Indeed, here, we consider the scenario where $f(N)=1$ which corresponds to an approximately linear increase of resources with $N$. We show below that despite this increase in resources, the subject's performance can decrease with set size as a result of the suboptimal use of these resources.

We also have to choose the prior distribution to be combined with the noise distribution {whose variance is} described in Equation~\ref{variance}. As in the rest of this paper, we can use the subject's prior $q(\mathbf{s})$ for this purpose. Alternatively, the experimenter's prior $p(\mathbf{s})$ might be used as the prior to be combined with the noise distribution.

To investigate the consequences of adaptation in the noise distribution for the precision-set size relationship, we compared four models: (i) a model with both suboptimal encoding and decoding (SED), where both the prior and the noise distribution (Equation~\ref{variance}) use the subject's prior $q(\mathbf{s})$; (ii) a suboptimal encoding (SE) model where the prior is identical to the experimenter's prior $p(\mathbf{s})$, but the noise distribution uses the subject's prior $q(\mathbf{s})$; (iii) a suboptimal decoding (SD) model where the prior is identical to the subject's prior, but the noise distribution uses the experimenter's prior and finally (iv) a uniform variance (UV) model where the prior is identical to the subject's prior and the noise variance is stimulus-independent. The UV model is identical to the models presented earlier in this paper. A subset of the combinations of encoding and decoding schemes above were used in a previous study within the context of a neural population coding model to investigate the consequences of sensory adaptation in stimulus encoding and/or decoding stages \cite{seriesetal2009}.

For the subject's prior, $q(\mathbf{s})$, either a Gaussian with uniform non-negative correlations or the mixture of Gaussians model is used. For the models with stimulus-dependent noise variance, the parameters $\sigma_0^2$ and $\delta$ were chosen such that (i) the performance of these models for set size 1 was approximately equal to the performance of the models with uniform noise variance considered previously in the paper and (ii) the stimulus-dependence of variance had an appreciable effect on model performance, because large values of $\delta$ diminish the effect of stimulus-dependence.

\subsubsection*{Memory precision-set size relationship}
\begin{figure}[!ht]
\centering
\includegraphics[scale=1.0]{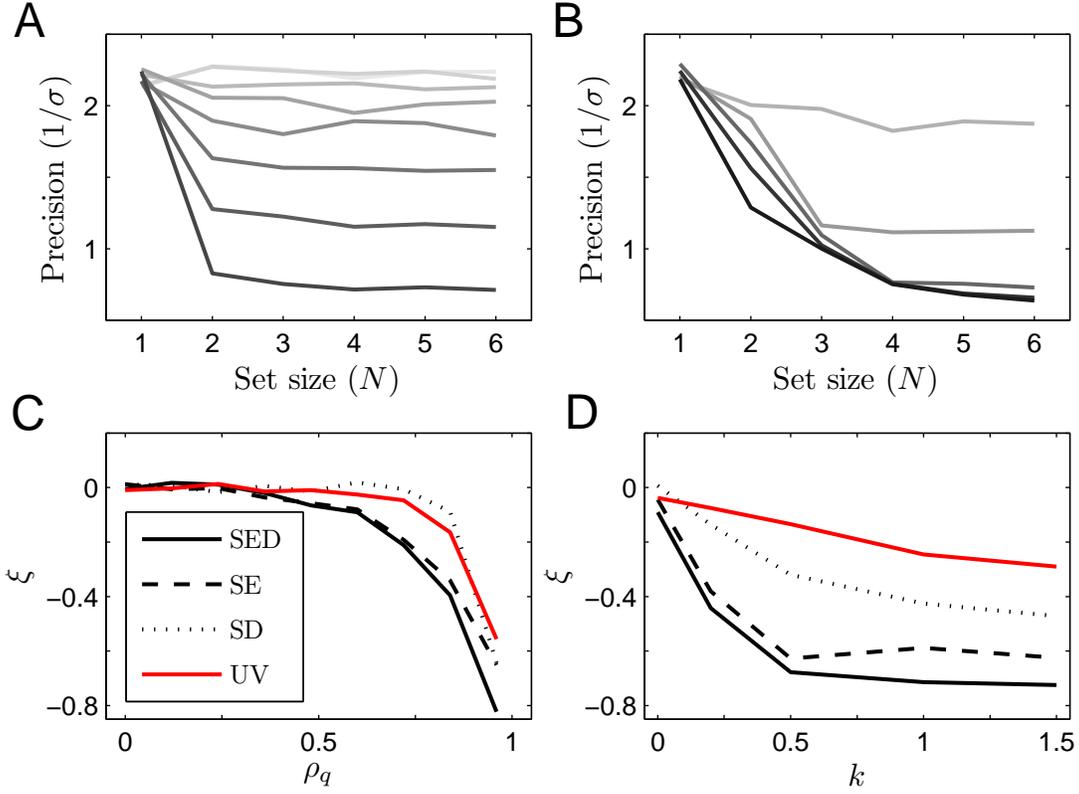} 
\caption{{\bf Simulation results for the case where $p(\mathbf{s})$ is an uncorrelated Gaussian (stimulus-dependent noise distribution).} (A-B) Precision-set size relationships for the SED model where $q(\mathbf{s})$ is either a multivariate Gaussian with uniform non-negative correlations $\rho_q$ (A) or a mixture of Gaussians (B). For the multivariate Gaussian model (A), 9 different $\rho_q$ values were tested ranging from 0 to 0.96 in increments of 0.12. Darker colors correspond larger $\rho_q$ values. For the mixture of Gaussians model (B), 5 different $k$ values were tested: $k=0,0.2,0.5,1,1.5$. Darker colors correspond to larger $k$ values. (C-D) Power-law exponents for different combinations encoding and decoding schemes where $q(\mathbf{s})$ is either a multivariate Gaussian with uniform non-negative correlations $\rho_q$ (C) or a mixture of Gaussians (D). Parameter values are $\sigma_0^2 = 0.025$ and $\delta = 0.01$. SED: suboptimal encoding and decoding, SE: suboptimal encoding, SD: suboptimal decoding, UV: uniform variance.}
\label{setsize_adap}
\end{figure}

Figure~\ref{setsize_adap} shows the precision-set size relationships predicted by the models when the experimenter's prior $p(\mathbf{s})$ is an uncorrelated Gaussian. Figure~\ref{setsize_adap}A-B shows the precision-set size curves for the SED model where $q(\mathbf{s})$ is a multivariate Gaussian with uniform non-negative correlations $\rho_q$ (Figure~\ref{setsize_adap}A) and a mixture of Gaussians (Figure~\ref{setsize_adap}B) respectively. Figure~\ref{setsize_adap}C-D shows the power-law exponents for different combinations of encoding and decoding schemes with either the multivariate Gaussian model (Figure~\ref{setsize_adap}C) or the mixture of Gaussians model (Figure~\ref{setsize_adap}D). 

Two prominent features of the results shown in Figure~\ref{setsize_adap} are worth highlighting. First, the models with suboptimal encoding (SE and SED) show the steepest declines in performance with set size (Figure~\ref{setsize_adap}C-D). This suggests that the models with uniform noise variance considered earlier in this paper might yield substantially larger set size effects (smaller $\xi$) when supplemented with a suboptimal encoding scheme without having to postulate a resource limitation. For example, in the case of multivariate Gaussian $q(\mathbf{s})$ (Figure~\ref{setsize_adap}C), the UV model produces a power-law exponent of $\xi = -0.16$ ($R^2 = 0.89$) for $\rho_q = 0.84$, whereas the SED model (which is obtained by replacing the stimulus-independent noise variance in the UV model with a stimulus-dependent noise variance as in Equation~\ref{variance}) has {$\xi = -0.40$ ($R^2 = 0.84$)} for the same $\rho_q$ value. In other words, with suboptimal encoding schemes, a smaller correlation in $q(\mathbf{s})$ is required to explain the steep declines in precision with set size.

It is important to emphasize that the models all assume that the resources increase approximately linearly with set size ($f(N) = 1$) and therefore the set size effects are caused entirely by suboptimal encoding and/or suboptimal decoding. This can also be verified by considering what happens when there is no model mismatch in the prior or in the noise distribution, that is, when $\rho_q=0$ in the SED model. In this case both the prior and the noise distribution use the experimenter's prior $p(\mathbf{s})$. The simulation results for this case are shown by the lightest line in Figure~\ref{setsize_adap}A. The precision-set size curve is approximately flat (the power-law exponent {$\xi=-0.01$}) suggesting that the model makes efficient use of the increase in resources with set size. The empirical distribution of precisions [distribution of $1/\sigma(\mathbf{s})$ computed from Equation~\ref{variance} where $\mathbf{s}$ is drawn from $p(\mathbf{s})$] over $10^5$ trials for each set size is shown in Figure~\ref{demo_stim_dependent_noise_fig}B. The distributions for different set sizes look slightly different, but they are mostly overlapping. Figure~\ref{demo_stim_dependent_noise_fig}C shows the empirical distribution of precisions over $10^5$ simulated trials when $p(\mathbf{s})$ is an uncorrelated Gaussian, but $q(\mathbf{s})$ is a multivariate Gaussian with uniform correlations ($\rho_q=0.9$). Note how the distributions shift to the left for larger set sizes. 

The second prominent feature of the results shown in Figure~\ref{setsize_adap} is the remarkably good power-law fits obtained with the suboptimal encoding schemes (SE and SED) using the mixture of Gaussians model with $k=1$ (or similar values) for $q(\mathbf{s})$ (Figure~\ref{setsize_adap}B). The SED model with $k=1$ yields $\xi=-0.71$ ($R^2=0.98$). This is a surprising result, because unlike in standard continuous resource models where the power-law relationship between precision and set size is explicitly built into the noise distribution, there is nothing in Equation~\ref{variance} that, on the face of it, suggests a power-law relationship between precision and set size. Indeed, the power-law fits are significantly worse when we use the multivariate Gaussian model with non-negative correlations for $q(\mathbf{s})$ (as can be visually verified by looking at the curves in Figure~\ref{setsize_adap}A). Thus, this appears to be a property of the mixture of Gaussians model (or similar models) specifically.  

\begin{figure}[!ht]
\centering
\includegraphics[scale=1.0]{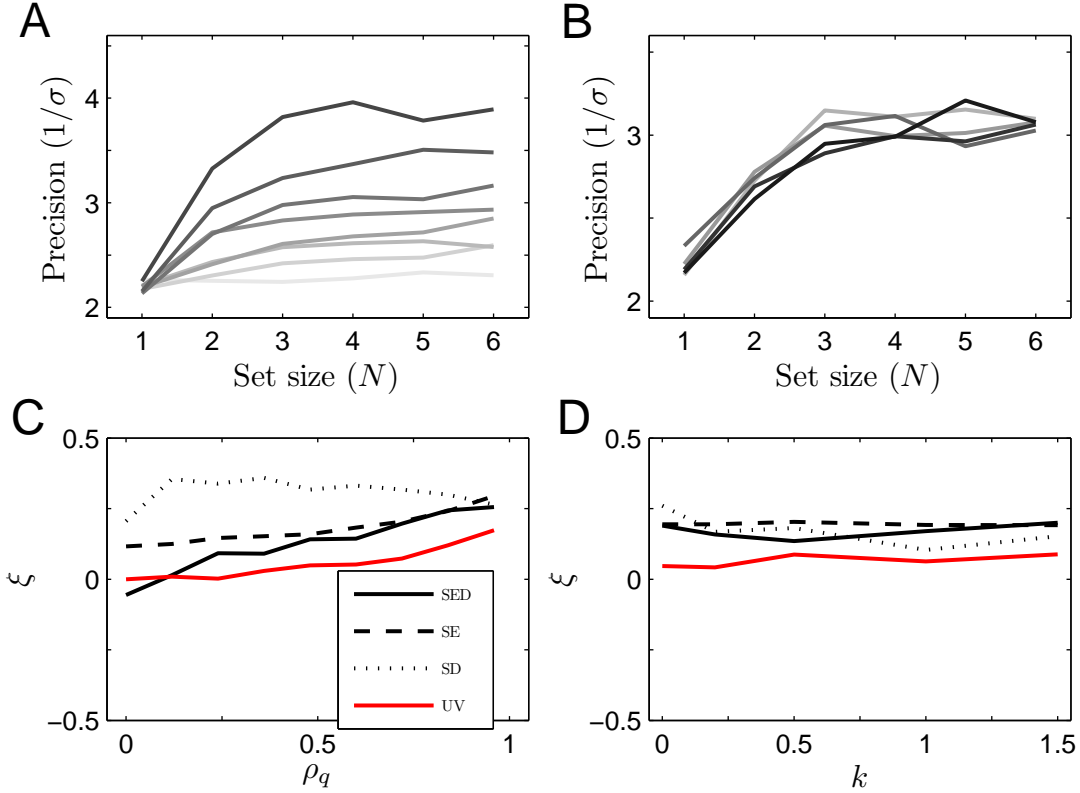} 
\caption{{\bf Simulation results for the case where $p(\mathbf{s})$ is correlated (stimulus-dependent noise distribution).} Similar to Figure~\ref{setsize_adap}, but the results are shown for the case where $p(\mathbf{s})$ is a multivariate Gaussian with uniform correlations ($\rho_p=0.96$).}
\label{setsize_adap_corr}
\end{figure}

Figure~\ref{setsize_adap_corr} shows the precision-set size relationships predicted by the models when the experimenter's prior $p(\mathbf{s})$ is a correlated Gaussian with $\rho_p = 0.96$. Inverse set size effects can be observed for all models. The SED models, which differ from the UV models only in the stimulus-dependence of the noise distribution, produce larger inverse set-size effects than the UV models. Figure~\ref{demo_stim_dependent_noise_fig}D shows the empirical distribution of precisions over $10^5$ simulated trials when both $p(\mathbf{s})$ and $q(\mathbf{s})$ are correlated Gaussians (with $\rho_q = \rho_p = 0.9$). Note how the distributions this time shift to the right for larger set sizes.  

\subsubsection*{Variability in {memory} precision}
An immediate consequence of using a stimulus-dependent noise variance as in Equation~\ref{variance} is variability in encoding precision {which automatically leads to variability in posterior precision}. With a stimulus-dependent noise variance, variability in encoding precision, {hence variability in memory precision}, arises even when we use a multivariate Gaussian model for $q(\mathbf{s})$. In fact, using any non-uniform distribution for $q(\mathbf{s})$ in Equation~\ref{variance} would lead to some variability in {memory} precision. Note that this is not the case if the noise variance is uniform. For example, if the noise variance is uniform, a multivariate Gaussian $q(\mathbf{s})$ does not lead to variability in {memory} precision as explained previously. {Also note that a stimulus-dependent noise variance in the noise distribution leads to encoding variability due to differences in noise precision for different stimulus configurations $\mathbf{s}$ (Equation~\ref{variance}). This is different from the purely stochastic, unstructured encoding variability hypothesized in \cite{vandenbergetal2012}, \cite{fougnieetal2012}}.

\begin{figure}[!ht]
\centering
\includegraphics[scale=1.0]{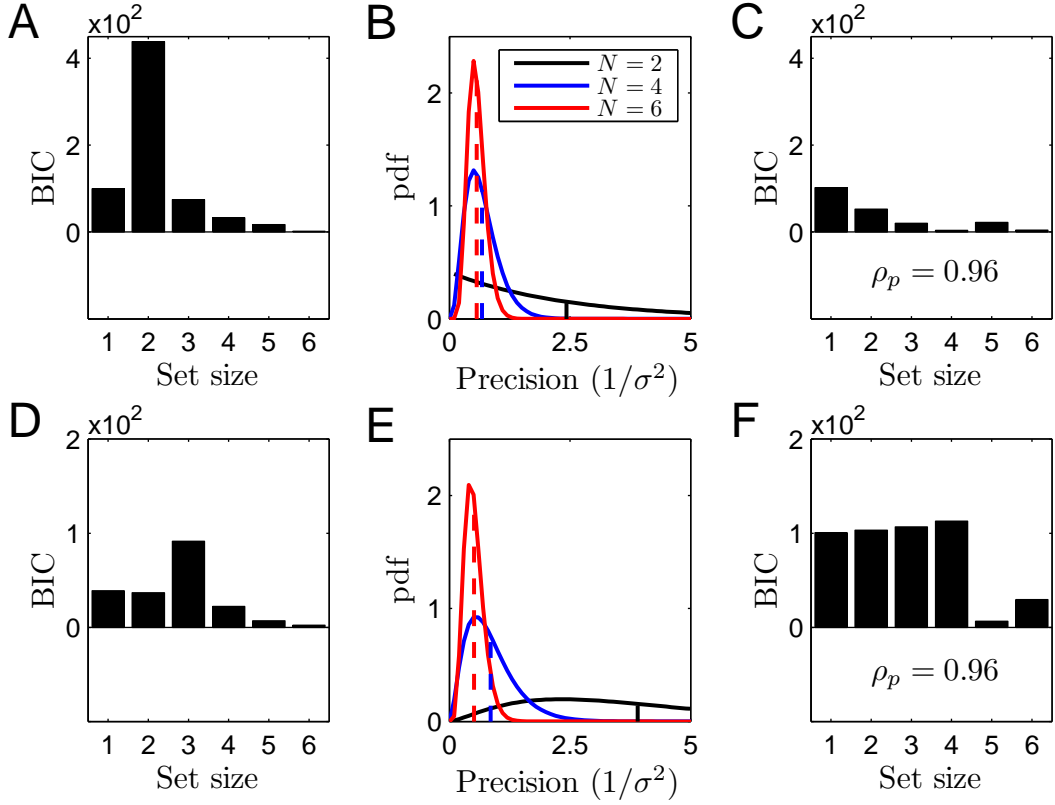} 
\caption{{\bf Variability in memory precision (stimulus-dependent noise distribution).} (A-B) Results for the case of uncorrelated Gaussian $p(\mathbf{s})$ and the SED model with Gaussian $q(\mathbf{s})$ with uniform non-negative correlations ($\rho_q = 0.96$): (A) BIC score of the Gaussian fit relative to the BIC score of the $t$-distribution fit for each set size. Lower scores indicate better fits. Positive values indicate better fits for the $t$-distribution than for the Gaussian distribution. (B) For three different set sizes, the estimated Gamma distributions over precision for the best fit $t$-distributions. The vertical lines show the means of the Gamma distributions. (C) Similar to (A), but here the experimenter's prior contains correlations ($\rho_p=0.96$). (D-E) Similar to (A-B) except the results are for the case of uncorrelated Gaussian $p(\mathbf{s})$ and the SED model with the mixture of Gaussians $q(\mathbf{s})$ ($k=1$). (F) Similar to (D), but the experimenter's prior contains correlations ($\rho_p=0.96$).}
\label{variability_adap}
\end{figure}

Figure~\ref{variability_adap} shows the variability in {memory} precision predicted by the SED model using a multivariate Gaussian distribution with $\rho_q = 0.96$ for $q(\mathbf{s})$ (Figure~\ref{variability_adap}A-C) or using the mixture of Gaussians model with $k=1$ (Figure~\ref{variability_adap}D-F). Compared to the uniform noise variance scenario considered earlier in the paper, three differences are apparent in this figure. First, variability in {memory} precision arises even when we use a multivariate Gaussian model for $q(\mathbf{s})$ (Figure~\ref{variability_adap}A-C), which was not the case with a uniform noise variance. Second, for both models, variability in {memory} precision arises even with set size 1. Again, none of the models with uniform noise variance yielded variability in {memory} precision for set size 1. A previous study found evidence for variability in {memory} precision even for set size 1 \cite{fougnieetal2012}, which suggests that the models with uniform noise variance considered earlier in this paper might not be able to explain the full range of variability in {memory} precision. Third, unlike in models with uniform noise variance, introducing correlations in the experimenter's prior did not have a drastic effect on variability in {\color{black}memory} precision for the SED models (Figure~\ref{variability_adap}C and F).

{\subsubsection*{Model mismatch does not necessarily imply large biases in the observer's responses}
A possible objection to the model mismatch account of performance limitations in VSTM is that without a resource limitation model mismatch requires strong correlations in the subject's prior to explain the magnitude of performance limitations typically observed in VSTM tasks. This in turn would imply strong biases toward the prior. Although such biases are observed in some studies \cite{wilkenma2004}, \cite{orhanjacobs2013}, \cite{bradyalvarez2011}, they are either weak or even absent in other studies. Moreover, even in studies where such biases are observed, it is unclear whether the size of the observed biases is large enough to be compatible with the biases that would be predicted by model mismatch. 

Here, we show that when adaptation in the noise distribution to natural stimulus statistics (i.e. mismatch in the noise distribution) is taken into account, model mismatch does not necessarily imply strong biases toward the prior. The crucial idea here is that a stimulus-dependent noise distribution with a variance of the form given in Equation~\ref{variance} generates biases \textit{away} from the prior. This point was first noted by Stocker and Simoncelli \cite{stockersimoncelli2005} in connection with the effects of visual adaptation on the observer's noise distribution and the explanation for the biases away from the prior here is the same as in their case. Intuitively, considered as a function of the stimulus $\mathbf{s}$, i.e. as a likelihood function, a Gaussian distribution with a stimulus-dependent variance of the form given in Equation~\ref{variance} is skewed away from the high probability regions of $q(\mathbf{s})$, because roughly speaking these regions have smaller variance and thus are ``narrower'' under the likelihood, whereas the low probability regions of $q(\mathbf{s})$ have high variance and thus are ``broader''. The negative biases (biases away from the prior) induced by the stimulus-dependent noise distribution counteract the positive biases (biases toward the prior) induced by the prior. Depending on the contribution of these two factors (the prior and the stimulus-dependent noise), it is possible to obtain different degrees of bias toward the prior. 

\begin{figure}[!ht]
\centering
\includegraphics[scale=1.0, trim=0mm 0mm 0mm 0.4mm,clip]{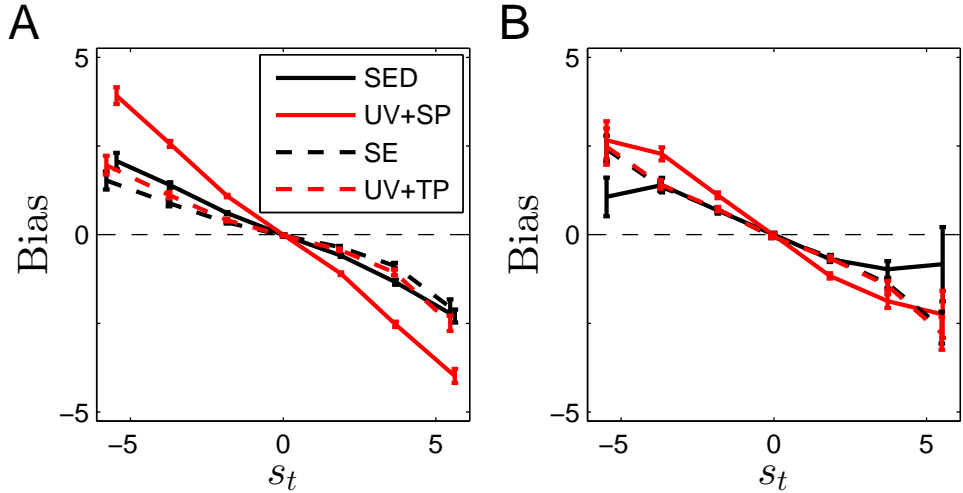} 
\caption{{\bf Comparison of the biases of various models.} Biases are computed by dividing the target stimulus value $s_t$ into 8 bins and measuring the mean and the standard error of the estimation errors, $\hat{s}_t - s_t$, for each bin. Results shown here are for $N=6$. (A) Subject's prior, $q(\mathbf{s})$, is a correlated Gaussian with uniform correlations ($\rho_q = 0.9$). (B) $q(\mathbf{s})$ is the mixture of Gaussians model ($k=1$). Other parameter values are: $\sigma_0^2 = 0.025$ and $\delta = 0.01$ (same as in previous simulations with stimulus-dependent noise distributions). SED: suboptimal encoding and decoding, UV+SP: uniform variance noise distribution combined with the subject's prior, SE: suboptimal encoding, UV+TP: uniform variance noise distribution combined with the true prior.}
\label{adaptation_in_noise_biases}
\end{figure}

In Figure~\ref{adaptation_in_noise_biases}, we compare the biases produced by four different models. The first model is the SED model ($\rho_q = 0.9$) with mismatched prior and noise distributions. The second model (UV+SP) is a uniform variance model with prior mismatch only ($\rho_q = 0.9$). Because the predicted biases strongly depend on the noise variance, to make a fair comparison with the SED model, the variance of the noise distribution in the UV+SP model is matched trial-by-trial to the variance of the SED model's noise distribution. The third model is the SE model ($\rho_q = 0.9$) with a mismatched noise distribution only (the prior used by the model is the true prior). The fourth model (UV+TP) is a uniform variance model with no prior mismatch. Again, for a fair comparison with the SE model, the variance of the noise distribution in the UV+TP model is matched trial-by-trial to the variance of the SE model's noise distribution. The crucial comparison is between the models with suboptimal encoding (SED and SE) and the UV+TP model. Because in our simulations even the true prior distribution is informative, biases predicted by the UV+TP model (which uses the true prior) should be taken as the baseline and all other biases should be compared to this baseline. 

Figure~\ref{adaptation_in_noise_biases}A shows that models with stimulus-dependent noise distributions (SED and SE) generate biases comparable to the baseline with the SED model producing slightly larger and the SE model producing slightly smaller biases than the baseline. As discussed above, this is due to the negative biases induced by the stimulus-dependent noise that counteract the positive biases induced by the prior. The only model that produces strong biases is the UV+SP model which incorporates only the prior mismatch. Qualitatively similar results are obtained when the mixture of Gaussians model is used for $q(\mathbf{s})$ instead of the Gaussian with uniform correlations (Figure~\ref{adaptation_in_noise_biases}B).

These results show that if the possibility of a mismatch in the noise distribution is taken into account, model mismatch does not necessarily imply large biases toward the prior.
}
\section*{Discussion}

In this paper, we presented a novel account of performance limitations in VSTM based on the idea that many of the main characteristics of these limitations can be explained, at least in part, as the result of a mismatch between the statistics of stimuli used in most VSTM experiments and the statistics of the same stimuli in the natural visual input that the subject's visual system might be adapted to. In particular, we showed that the decline in memory precision with increasing set size, variability in encoding precision and the different set size dependencies of initial encoding rate and asymptotic precision can all be explained, in principle, from this perspective without assuming any resource limitation. 

It is ultimately an empirical question to what extent model mismatch and a fixed resource limitation might be contributing to performance limitations in standard VSTM studies. {This question should be addressed experimentally using sufficiently powerful experimental paradigms and computational models that can capture the effects of both model mismatch (in the form of prior mismatch and/or mismatch in the noise distribution) and a potential resource limitation.}

There is already some experimental evidence suggesting that model mismatch plays at least \textit{some} role in many VSTM studies. For example, subjects' responses display biases \cite{wilkenma2004}, \cite{bradyalvarez2011}, \cite{huangsekuler2010}, and dependencies \cite{orhanjacobs2013}, \cite{jiangetal2000} and, on a trial by trial basis, their responses tend to be more accurate when, by chance, the presented visual display contains regularities (e.g., homogeneity) in a given trial even when the stimuli are drawn independently and uniformly across trials \cite{orhanjacobs2013}, \cite{bradyalvarez2012}, \cite{bradytenenbaum2013}. These results would be difficult to explain if subjects were using an internal model that exactly matched the generative model used by the experimenter to generate the stimuli. To the extent that model mismatch plays \textit{some} role in VSTM tasks, we believe that the role of capacity or resource limitations in VSTM has been overstated by researchers.

Consistent with the predictions of our framework, it has been observed that subjects can hold more items in VSTM and with better precision if the stimuli are more homogeneous \cite{bradytenenbaum2013}, \cite{linluck2009}, \cite{simsetal2012}. But our results suggest that homogeneity is not the only property that would reduce, eliminate or reverse the set size effects. In general, any manipulation that would bring the stimulus statistics used in the experiment closer to the environmental statistics, or to the subject's prior, would be expected to boost the subject's performance and to reduce or even reverse the set size effects. For example, using oriented patches that form smooth contours as stimuli rather than random configurations of orientations would also be predicted to reduce these effects. Our framework also crucially predicts these same effects on a trial-by-trial basis---even when the experimenter's prior is an independent and uniform distribution, by chance, stimuli will be more or less natural in different trials according to the subject's prior. Subjects are predicted to perform better in more natural trials than in less natural trials as has already been demonstrated experimentally \cite{orhanjacobs2013}, \cite{bradyalvarez2012}, \cite{bradytenenbaum2013}. {It is also interesting to note that effects similar to the inverse set-size effect, a crucial prediction of our framework when the experimenter's prior matches the subject's prior, where it becomes easier to recall or recognize a feature or object within the context of other correlated features or objects than in isolation, have been reported previously with ecologically relevant stimuli such as words and faces \cite{reicher1969}, \cite{tanakafarah1993}.} 

The role of model mismatch, and of suboptimal inference in general, in behavioral performance has been emphasized recently by Beck and colleagues \cite{becketal2012}. Beck et al. \cite{becketal2012} note the tendency in systems neuroscience to see neural variability as the primary culprit in behavioral variability. Neural variability is closely related to the concept of internal noise used in signal detection theory and in ideal observer models of many standard psychophysical tasks \cite{ma2010}, including VSTM tasks \cite{wilkenma2004}, \cite{mahuang2009}, and there is a similar tendency in these models to assume perfect knowledge of the task (including perfect knowledge of the statistics of the stimuli used in the task) and to attribute behavioral imperfections to internal noise. Beck et al. \cite{becketal2012} show instead that having a suboptimal model of the task is likely to be a major contributor to behavioral variability, especially for complex, real-world tasks, as neural variability can always be averaged out, but a suboptimal model cannot. Our simulation results indeed suggest that the effects of model mismatch become more severe in environments with strong statistical regularities, which is presumably the ecologically relevant stimulus regime (Figure~\ref{strong_corr_demo}).

Our results highlight the importance of the stimulus distributions used in experimental studies. Ecologically realistic stimulus distributions that more closely match the internal models of subjects can lead to qualitatively very different patterns of behavior than ecologically unrealistic stimulus distributions. This suggests that researchers should be cautious about generalizing the results of experiments using ecologically unrealistic stimuli to ecologically more realistic settings. The potential problems with generalizations of this sort have long been known to experimental psychologists \cite{brunswik1943}, \cite{brunswik1955}, \cite{neisser1976}. 

In theoretical neuroscience, the idea that a memory system's storage and retrieval performance should depend on the stimulus statistics is not a new insight. It is known, for instance, that systems that appear to have restrictive memory capacity limitations for a certain type of stimulus statistic can operate in an entirely different and much less restrictive regime if a different kind of stimulus statistic is assumed, such as correlated vs.\ uncorrelated \cite{tsodyksfeigelman1988} or sparse vs.\ non-sparse stimuli \cite{gangulietal2008}, \cite{gangulisompolinsky2010}.       

It could be argued that our framework is unjustifiably restrictive in assuming that subjects cannot override their natural internal models and successfully adapt to stimulus statistics used in an experiment. It is important to emphasize that we do not claim that subjects cannot adapt to any aspect of the experimenter's prior. Some aspects of the experimenter's prior, such as the stimulus range or the mean, might be easy for subjects to learn during the course of an experiment. However, it is widely documented that subjects cannot always learn a statistical regularity, or the lack thereof, in an experimental task \cite{newportaslin2004}, \cite{creeletal2004}, \cite{fiseraslin2005}, \cite{micheljacobs2007}, \cite{seydelletal2010}. In general, subjects tend to learn statistical regularities with greater ease and success when these regularities are more ``natural'' (i.e., when they are consistent with the statistical regularities exhibited in the natural environment). Learning ``unnatural'' regularities requires a much longer training period \cite{schwarzkopfkourtzi2008} and tends to be highly stimulus-specific \cite{zhangkourtzi2010}. Similarly, subjects sometimes make inaccurate assumptions about the statistical structure of the stimuli used in a task. For example, they may assume dependencies between certain latent variables in the task \cite{turnhametal2011}, between stimuli in a given trial \cite{orhanjacobs2013} or between stimuli across trials \cite{huangsekuler2010}, \cite{yucohen2009} when, in fact, there are no such dependencies. These results suggest that in many, perhaps most, behavioral experiments, there is some discrepancy between the actual stimulus statistics used in the experiment and the subject's internal model. The extent of this discrepancy depends on the particular behavioral task as well as the actual stimulus statistics used in the experiment. Most of the simulations reported in this paper used a minimal form of discrepancy between the experimenter's prior and the subject's prior. In particular, we assumed only a mismatch between the correlation structures of the respective priors and did not assume any mismatches between the means or the marginal variances of the two distributions. 

Our framework can be extended to performance limitations in other domains. In multiple-object tracking, performance limitations similar to those in VSTM have been observed \cite{pylyshynstorm1988}. As in VSTM, these limitations have been explained in terms of a limited amount of attentional resources making subjects' representations noisier as the number of trajectories to be tracked is increased \cite{alvarezfranconeri2007}, \cite{vuletal2009}. Again, we hypothesize that at least part of these limitations can be explained in terms of a mismatch between the statistics of motion patterns generally used in multiple-object tracking experiments and the environmental statistics of the same variables that the subjects' visual system would be expected to be adapted to. Indeed, it is known that, for a given set size (i.e., number of trajectories to be tracked), motion trajectories that display certain non-trivial properties are easier to track for subjects than trajectories that do not have those properties \cite{yantis1992}. According to our framework, these types of trajectories can be considered as more likely trajectories under the subjects' prior, which may, in turn, have a basis in the environmental statistics of motion trajectories. 

Special populations display significant performance differences in VSTM tasks from normal adults. For example, children perform significantly worse than adults in VSTM tasks even when factors unrelated to VSTM are controlled for \cite{gathercoleetal2004}, \cite{burnettheyesetal2012}. Performance in VSTM tasks tends to be impaired in schizophrenia and some other mental illnesses \cite{farmeretal2000}, \cite{fulleretal2005}. Moreover, there are significant individual differences in VSTM performance among the general population \cite{fukudaetal2010}. These differences are generally explained in terms of differences in capacity (i.e., differences in the amount of attentional, neural or cognitive resources). Our framework offers a complementary account of performance differences in VSTM tasks. Our results show that differences in subjects' internal models for the experimental stimuli will manifest themselves as differences in their performance in these VSTM tasks. Thus, for example, individual differences in VSTM performance among the general population may be at least partly due to each individual's slightly different internal model of the experimental stimuli or to differences in their ability to adapt these internal models to different stimulus statistics, in addition to any differences in attentional, neural or cognitive resources that individuals may have. Similarly, individuals with schizophrenia may be relying on prior models substantially different from the priors used by normal adults in VSTM tasks. Consistent with this view, there is evidence suggesting that perceptual organization abilities are significantly impaired in individuals with schizophrenia \cite{silversteinetal2000}.  

Our results raise the possibility that many phenomena in the VSTM literature attributed to intrinsic properties of the VSTM system may in fact be explained, at least in part, as straightforward consequences of the prior expectations and/or the internal noise properties of the visual system due to adaptation to the regularities in the natural visual environment and the mismatch between those prior expectations and internal noise properties and the stimulus statistics used in many VSTM tasks. We demonstrated that in many cases it may be unnecessary to postulate bottlenecks in visual attention or neural variability or other sophisticated mechanisms to account for characteristics of subjects' performances in VSTM tasks. These characteristics may fall out naturally from very general principles governing an organism's adaptation to its natural visual environment, and probing that organism's visual system with unnatural stimulus statistics that it is not adapted to.

\section*{Acknowledgments}
We thank Chris Sims and Dave Knill for helpful discussions. This work was supported by research grants from the National Science Foundation (DRL-0817250) and the Air Force Office of Scientific Research (FA9550-12-1-0303) awarded to Robert Jacobs.

\bibliographystyle{unsrt}
\bibliography{OrhanJacobs2014}
\newpage

\section*{Supporting Information}
\renewcommand\thefigure{S\arabic{figure}}
\setcounter{figure}{0}    
\renewcommand\theequation{S\arabic{equation}}
\setcounter{equation}{0}    
  
\subsection*{Text S1. Simulation details}
\label{app_ch3_b}
Because the prior mean is always equal to a vector of zeros, for the multivariate Gaussian model with uniform correlations and the multivariate Gaussian model with random positive correlations, the posterior means are found by: $C\Sigma^{-1}\mathbf{x}$, where $C$ is the posterior covariance matrix, $\Sigma$ is the covariance matrix of the noise distribution and $\mathbf{x}$ is the noisy observation vector in a particular trial. The model's estimate for the target stimulus value is then equal to the corresponding element in the posterior mean vector. 

For the multivariate Gaussian model with random, positive correlations, we use the \texttt{gallery} function in Matlab with the \texttt{`randcorr'} option to generate random correlation matrices. Eigenvalues used as seeds for the \texttt{gallery} function are generated by first randomly drawing a sample from a symmetric Dirichlet distribution with concentration parameter $\gamma$ and multiplying it by the number of eigenvalues $N$ (this ensures that the eigenvalues sum to $N$). To prevent numerical instabilities, a small amount of independent positive random noise is added to each eigenvalue (uniform between 0 and $0.0001\times N$) and then the eigenvalues are re-normalized to sum to $N$. The \texttt{gallery} function generates correlation matrices with both negative and positive entries. To obtain correlation matrices with positive entries only, we simply take the absolute value of each entry in the correlation matrix. Although this procedure occasionally changes the eigenvalues of the correlation matrix, we confirmed, through simulations, that the resulting changes did not significantly alter the qualitative characteristics of the eigenvalue distribution of the correlation matrices. In particular, small $\gamma$ values still yield sparse eigenvalue distributions, whereas large $\gamma$ values yield broad eigenvalue distributions.

For the mixture of Gaussians model, the joint and marginal posterior means can be calculated analytically. We first derive the joint posterior $p(\mathbf{s}|\mathbf{x})$:
\begin{eqnarray}
p(\mathbf{s}|\mathbf{x}) & = & \frac{p(\mathbf{x}|\mathbf{s})p(\mathbf{s})}{p(\mathbf{x})} \\
& = & \frac{p(\mathbf{x}|\mathbf{s}) [\sum_{k=1}^N w_k p_k(\mathbf{s})]}{p(\mathbf{x})} 
\end{eqnarray}
where $p_k(\mathbf{s})$ is the $k^{\rm th}$ normal component in the mixture, with mean $\mathbf{0}$, covariance $\Gamma_k$ and weight $w_k$. Then:
\begin{eqnarray}
p(\mathbf{s}|\mathbf{x}) & = & \frac{\sum_{k=1}^N w_k p(\mathbf{x}|\mathbf{s}) p_k(\mathbf{s})}{p(\mathbf{x})}
\end{eqnarray}
In the last equation, $p(\mathbf{x}|\mathbf{s})$ and $p_k(\mathbf{s})$ are both Gaussians. The product of two Gaussians is another Gaussian, but no longer normalized. Therefore, the last equation can be re-written as: 
\begin{eqnarray}
p(\mathbf{s}|\mathbf{x}) & = & \frac{\sum_{k=1}^N w_k Z_k \mathcal{N}(\mathbf{s}; \mathbf{m}_k, C_k)}{p(\mathbf{x})} 
\label{appendix_posterior}
\end{eqnarray}
where $C_k = (\Sigma^{-1}+\Gamma_k^{-1})^{-1}$ is the posterior covariance matrix of the $k^{\rm th}$ component ($\Sigma$ being the covariance matrix of the noise distribution), $\mathbf{m}_k = C_k \Sigma^{-1}\mathbf{x}$ is the posterior mean of the $k^{\rm th}$ component (note that the prior means of all components are equal to $\mathbf{0}$) and $Z_k$ is the normalization constant for the $k^{\rm th}$ component:
\begin{eqnarray}
Z_k = (2\pi)^{-N/2} |C_k|^{1/2} |\Sigma|^{-1/2} |\Gamma_k|^{-1/2} \exp [-\frac{1}{2} (\mathbf{x}^T \Sigma^{-1} \mathbf{x} - \mathbf{m}_k^T C_k^{-1}\mathbf{m}_k)] 
\end{eqnarray}
In this normalization constant, we can ignore factors common to all components, because they cancel out during normalization. Thus, Equation~\ref{appendix_posterior} can be re-written as:
\begin{equation}
p(\mathbf{s}|\mathbf{x}) \propto \sum_{k=1}^N \tilde{w}_k \mathcal{N}(\mathbf{s}; \mathbf{m}_k, C_k)
\label{appendix_posterior_2}
\end{equation}
where $\tilde{w}_k = w_k |C_k|^{1/2} |\Gamma_k|^{-1/2} \exp (\frac{1}{2} \mathbf{m}_k^T C_k^{-1}\mathbf{m}_k)$ is the posterior weight of the $k^{\rm th}$ component. Joint and marginal posterior means can then be easily calculated from Equation~\ref{appendix_posterior_2} as weighted averages of joint and marginal means of individual components.

The default unnormalized prior mixture weights are $10^{16}$, $5\times 10^{15}$, $10^0$, $10^{-5}$, $10^{-10}$, $10^{-15}$ from the most homogeneous to the least homogeneous components, respectively.

For the models with stimulus-dependent noise variance presented in the main text, posterior means or modes are no longer analytically tractable. For these models, we used a Metropolis-Hastings algorithm to sample from the joint posterior. The proposal distribution was an isotropic Gaussian with unit variance  along each dimension. The algorithm was implemented using the \texttt{mhsample} function in Matlab. In each trial, 500 samples were drawn from the posterior. We used the mean of the marginal posterior as the model's estimate of the target stimulus value in a given trial. 2500 trials per set size were simulated in simulations with the multivariate Gaussian $q(\mathbf{s})$ with uniform non-negative correlations, and 1000 trials per set size in simulations with the mixture of Gaussians model. For the mixture of Gaussians model, the unnormalized prior mixture weights were $10^{16}$, $10^{15}$, $10^0$, $10^{-5}$, $10^{-10}$, $10^{-15}$ from the most homogeneous to the least homogeneous components, respectively. Parameter values for the stimulus-dependent noise distribution were as follows: $\sigma_0^2 = 0.025$ and $\delta = 0.01$.
 
Matlab code for running all the simulations and generating the figures reported in this paper can be obtained from \texttt{http://www.cns.nyu.edu/$\sim$eorhan/}.

\subsection*{Text S2. Fitting normal and $t$ distributions to the models' errors}
\label{app_ch3_c}
To test for variability in encoding precision, we fit normal and $t$ distributions to the models' errors. For this purpose, we use the \texttt{`fitdist'} function in Matlab's Statistics Toolbox with the \texttt{`normal'} and \texttt{`tlocationscale'} options respectively. BIC scores of the two distributions are then computed according to: $BIC = -2 \log L + k \log n$, where $\log L$ is the log-likelihood of the model evaluated at the estimated parameter values, $n$ is the number of data points and $k$ is the number of free parameters. The normal distribution has two free parameters (mean and variance parameters), whereas the $t$ distribution has three free parameters (location $\mu$, scale $\sigma$ and degrees-of-freedom parameter $\nu$). Given the estimated parameters of the $t$ distribution, $\hat{\mu}$, $\hat{\sigma}$ and $\hat{\nu}$, the parameters of the implied Gamma distribution over precisions can be computed according to: $\hat{\alpha} = \frac{\hat{\nu}}{2}$ and $\hat{\beta} = \frac{\hat{\sigma}^2 \hat{\nu}}{2}$.

\subsection*{Text S3. Derivation of the precision of the error distribution for the Gaussian model with uniform correlations}
\begin{figure}
\centering
\includegraphics[scale=1]{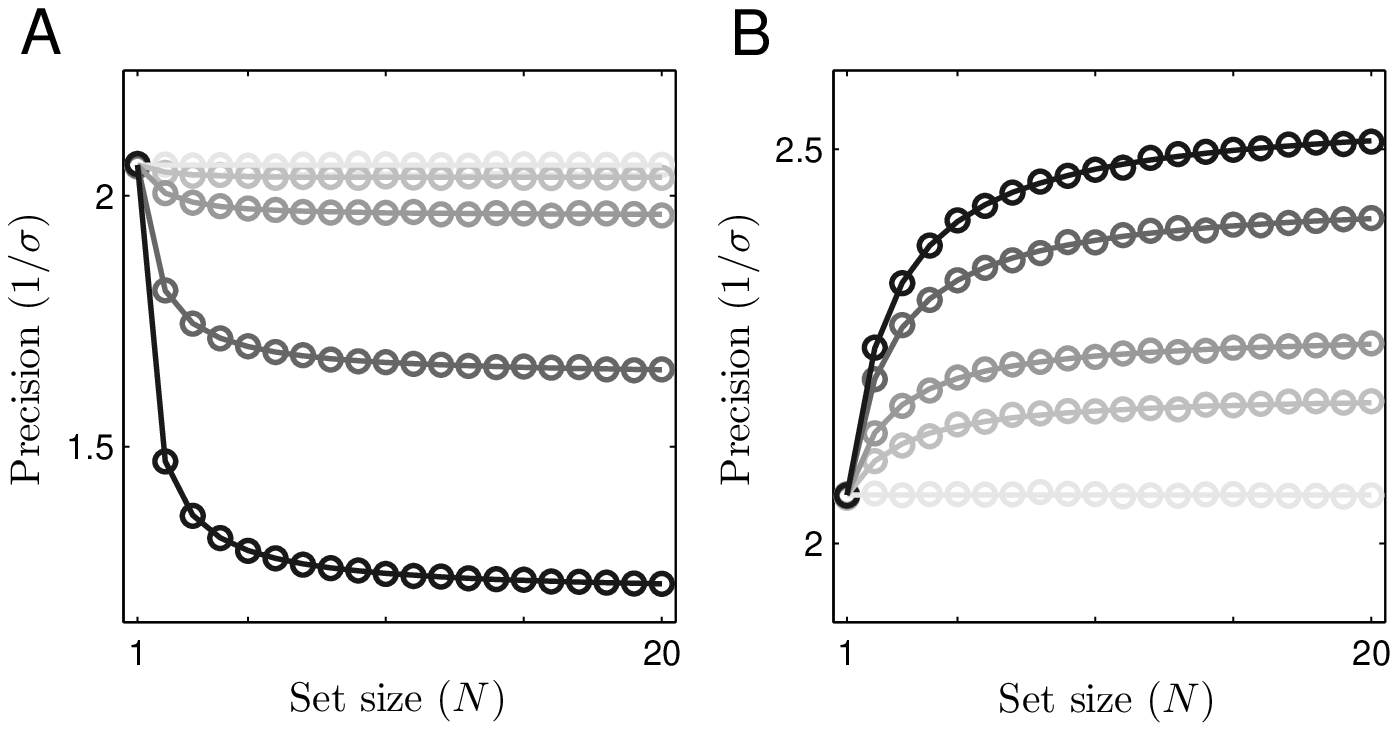} 
\caption{\textbf{Comparison of analytic solution (solid lines) with simulation results (circles).} (A) $p(\mathbf{s})$ has no correlations ($t=0$); (B) correlated $p(\mathbf{s})$ with $t=0.9$. Darker colors correspond to larger $r$ values ($r=0,0.4, 0.6, 0.8, 0.9$ respectively). Other parameter values are as follows: $\sigma^2 = 0.25$, $p^2=q^2=4$.}
\label{analytic_simulation}
\end{figure}

The covariance matrix of the posterior is given by $C = (\Sigma^{-1} + Q^{-1})^{-1}$ where $\Sigma^{-1} = (1/\sigma^2)I$ is the precision matrix of the noise distribution and $Q^{-1}$ is the precision matrix of the prior. It can be shown that $Q^{-1}$ has the form:
\begin{equation}
Q^{-1} = \frac{1}{q^2} \begin{bmatrix}
       \alpha &        & -\beta           \\
              & \ddots &                  \\
       -\beta &        & \alpha
     \end{bmatrix}
\end{equation} 
where $\alpha = \frac{(N-2)r + 1}{(1-r)[(N-1)r + 1]}$, $\beta = \frac{r}{(1-r)[(N-1)r + 1]}$, $q^2$ is the marginal variance of the prior, $\sigma^2$ is the marginal variance of the noise distribution and $r$ is the correlation coefficient of the prior. Then:
\begin{equation}
\Sigma^{-1} + Q^{-1} = \frac{1}{q^2} \begin{bmatrix}
       \alpha^{*} &        & -\beta           \\
              & \ddots &                  \\
       -\beta &        & \alpha^{*}
     								   \end{bmatrix} = \frac{\alpha^{*}}{q^2} \begin{bmatrix}
     								          1 &        & -\beta/\alpha^{*}          \\
     								                 & \ddots &                  \\
     								          -\beta/\alpha^{*} &        & 1
     								        								   \end{bmatrix}
\end{equation}
where $\alpha^{*} = \alpha + \frac{q^2}{\sigma^2}$. We can now find the posterior covariance, $C$:
\begin{equation}
C = (\Sigma^{-1} + Q^{-1})^{-1} = \frac{q^2}{\alpha^{*}} \begin{bmatrix}
       \alpha^{**} &        & \beta^{**}           \\
              & \ddots &                  \\
       \beta^{**} &        & \alpha^{**}
     								   \end{bmatrix}
\end{equation}
with:
\begin{equation}
\alpha^{**} = \frac{(N-2)(-\frac{\beta}{\alpha^{*}})+1}{(1+\frac{\beta}{\alpha^{*}})[(N-1)(-\frac{\beta}{\alpha^{*}})+1]}
\end{equation}
and
\begin{equation}
\beta^{**} = \frac{\frac{\beta}{\alpha^{*}}}{(1 + \frac{\beta}{\alpha^{*}})[(N-1)(-\frac{\beta}{\alpha^{*}})+1]}
\end{equation}
We would like to find out the marginal variance of the error distribution $p(\hat{\mathbf{s}} - \mathbf{s})$. Let us denote $\mathbf{e} = \hat{\mathbf{s}} - \mathbf{s}$. We first note that: 
\begin{equation}
p(\hat{\mathbf{s}}|\mathbf{s}) = \int p(\hat{\mathbf{s}}|\mathbf{x})p(\mathbf{x}|\mathbf{s})d\mathbf{s} = \int \delta(\hat{\mathbf{s}}=C\Sigma^{-1}\mathbf{x}) \mathcal{N}(\mathbf{x}; \mathbf{s}, \Sigma)d\mathbf{s} = \mathcal{N}(\hat{\mathbf{s}}; C\Sigma^{-1}\mathbf{s}, C\Sigma^{-1}C)
\end{equation}
Note that $C$ and $\Sigma^{-1}$ are symmetric matrices, thus they are identical to their transposes. Then, $p(\mathbf{e}|\mathbf{s}) = \mathcal{N}(\mathbf{e}; (C\Sigma^{-1} - I)\mathbf{s}, C\Sigma^{-1}C)$ and the error distribution is given by: 
\begin{eqnarray}
p(\mathbf{e}) & = & \int p(\mathbf{e}|\mathbf{s}) p(\mathbf{s}) d\mathbf{s} = \int  \mathcal{N}(\mathbf{e}; (C\Sigma^{-1} - I)\mathbf{s}, C\Sigma^{-1}C) \mathcal{N}(\mathbf{s}; \mathbf{0},P) d\mathbf{s} \\
 & = & \mathcal{N}(\mathbf{e}; \mathbf{0}, (C\Sigma^{-1}-I)P(C\Sigma^{-1} - I)^T + C\Sigma^{-1}C) \label{big_equation}
\end{eqnarray}
where $P$ is the covariance matrix of the experimenter's prior, $p(\mathbf{s})$. We assume a covariance matrix of the form: 
\begin{equation}
P = \begin{bmatrix}
       p^2 &        & t p^2  \\
            & \ddots &        \\
       t p^2 &        & p^2
    \end{bmatrix}
\end{equation}
with marginal variance $p^2$ and correlation coefficient $t$. We are only interested in a single marginal variance of the error distribution (all marginal variances are identical due to symmetry). Thus, without loss of generality, we can concentrate on a single diagonal entry in the covariance matrix in Equation~\ref{big_equation}. We start by noting: 
\begin{equation}
C\Sigma^{-1} - I = \frac{q^2}{\sigma^2 \alpha^{*}} \begin{bmatrix}
       \gamma     &        & \beta^{**}           \\
                  & \ddots &                  \\
       \beta^{**} &        & \gamma
     								   \end{bmatrix}
\end{equation}
where $\gamma = \alpha^{**} - \frac{\sigma^2\alpha^{*}}{q^2}$. After a tedious but straightforward derivation, it can be shown that: 
\begin{equation}
[(C\Sigma^{-1}-I)P(C\Sigma^{-1} - I)^T]_{ii} = (\frac{q^2 p }{\sigma^2 \alpha^{*}})^2\left[\gamma^2 + 2(N-1)\gamma \beta^{**}t + (N-1)[{\beta^{**}}^2 + (N-2){\beta^{**}}^2 t]\right] \label{part_1}
\end{equation}
Similarly, a straightforward derivation leads to: 
\begin{equation}
[C\Sigma^{-1}C]_{ii} = [\frac{1}{\sigma^2}CC]_{ii} = (\frac{q^4}{\sigma^2 {\alpha^{*}}^2})\left[{\alpha^{*}}^2 + (N-1){\beta^{**}}^2 \right] \label{part_2}
\end{equation}
The marginal variance of the error distribution is obtained by adding Equations~\ref{part_1} and \ref{part_2}. Figure~\ref{analytic_simulation} compares the analytic solution with simulation results for set sizes up to $N=20$. This figure shows that simulation results obtained over $10^6$ simulated trials closely match the analytic predictions. 

\begin{figure}
\centering
\includegraphics[scale=1]{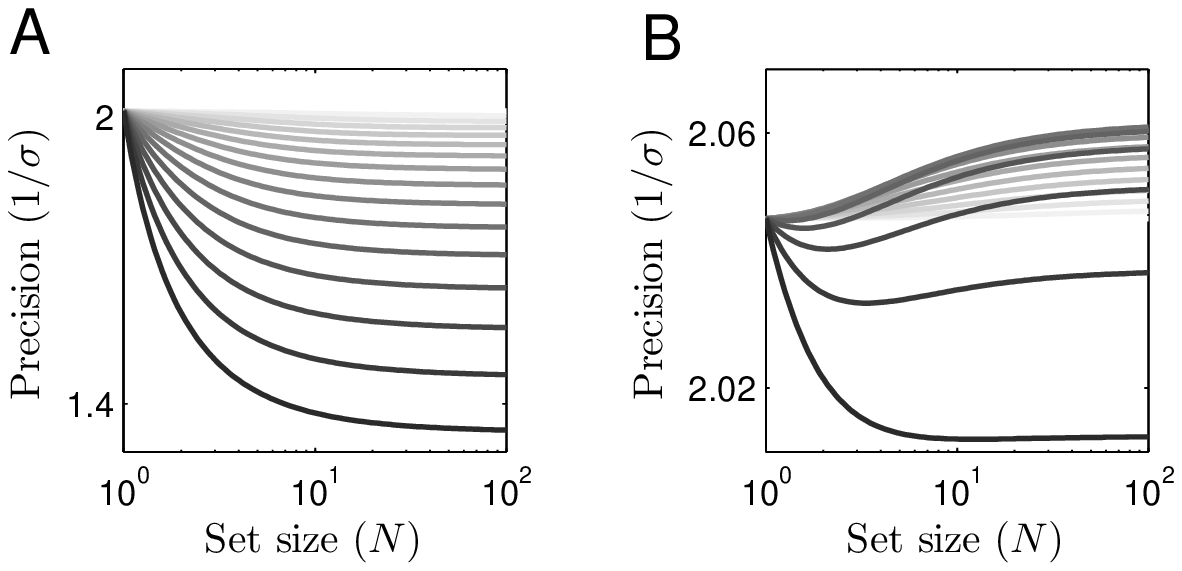} 
\caption{\textbf{The effects of underestimating vs. overestimating the variance.} (A) The variance is underestimated ($p^2=4$ and $q^2=2$). (B) The variance is overestimated ($p^2=4$ and $q^2=8$). In both cases, $p(\mathbf{s})$ is uncorrelated ($t=0$) and different lines correspond to different correlation values in the subject's prior ($r$ ranges from 0 to 0.75 in steps of 0.05 with darker colors indicating larger values). The noise variance $\sigma^2 = 0.25$ in both cases. Note the logarithmic scale in the \textit{x}-axis.}
\label{under_overestim}
\end{figure}

Using the analytic expression for the precision of the error distribution, Figure~\ref{under_overestim} shows that underestimating (Figure~\ref{under_overestim}A) and overestimating (Figure~\ref{under_overestim}B) the marginal variance of the actual prior have different effects on the precision of the error distribution. In general, underestimating the actual variance has a more detrimental effect on precision than overestimating it. Figure~\ref{under_overestim}B shows that when the variance is underestimated, it may actually be beneficial to introduce correlations in the subject's prior. Introducing moderate correlations can also lead to an inverse set size effect where precision increases with set size (for the range of set sizes tested here), even if the experimenter's prior $p(\mathbf{s})$ is uncorrelated (Figure~\ref{under_overestim}B). However, the magnitude of this inverse set size effect is rather small.

\begin{figure}
\centering
\includegraphics[scale=1]{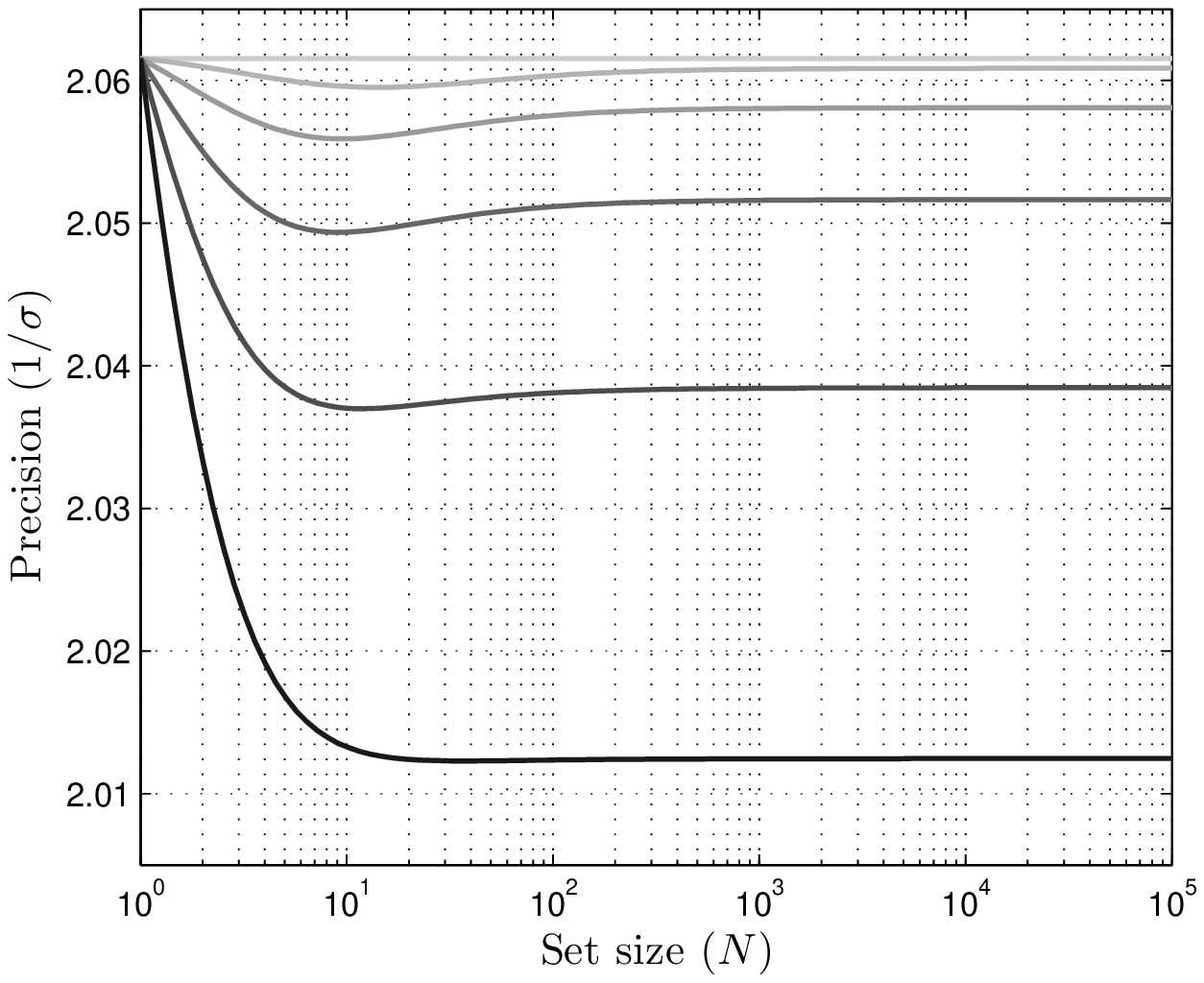} 
\caption{\textbf{Precision-set size curves for 6 different $r$ values (from 0 to 0.5 in steps of 0.1) when $p^2=q^2=4$ and the experimenter's prior is uncorrelated ($t=0$).} Darker colors correspond to larger $r$ values. The noise variance, $\sigma^2$, is 0.25. Note the logarithmic scale in the \textit{x}-axis.}
\label{nonmonotonic}
\end{figure}

The analytic expression for the precision of the error distribution allows us to calculate precision for very large set sizes. Surprisingly, even when the marginal variance is correctly estimated ($p^2=q^2$), and thus only the correlation is misestimated ($t \neq r$), precision is not always a monotonically decreasing function of $N$ (Figure~\ref{nonmonotonic}). Figure~\ref{nonmonotonic} shows the precision-set size relationship for 6 different $r$ values, when $p^2=q^2=4$ and the experimenter's prior is uncorrelated ($t=0$). Precision is generally a decreasing function of $N$ for set sizes up to $N\approx 10$; it slightly increases for larger set sizes and then asymptotes. The slight increase in precision for set sizes larger than $N \approx 10$ becomes less prominent with larger $r$ values. 

\subsection*{Text S4. Biases and dependencies in model responses}

\begin{figure}[t!]
\centering
\includegraphics[scale=1]{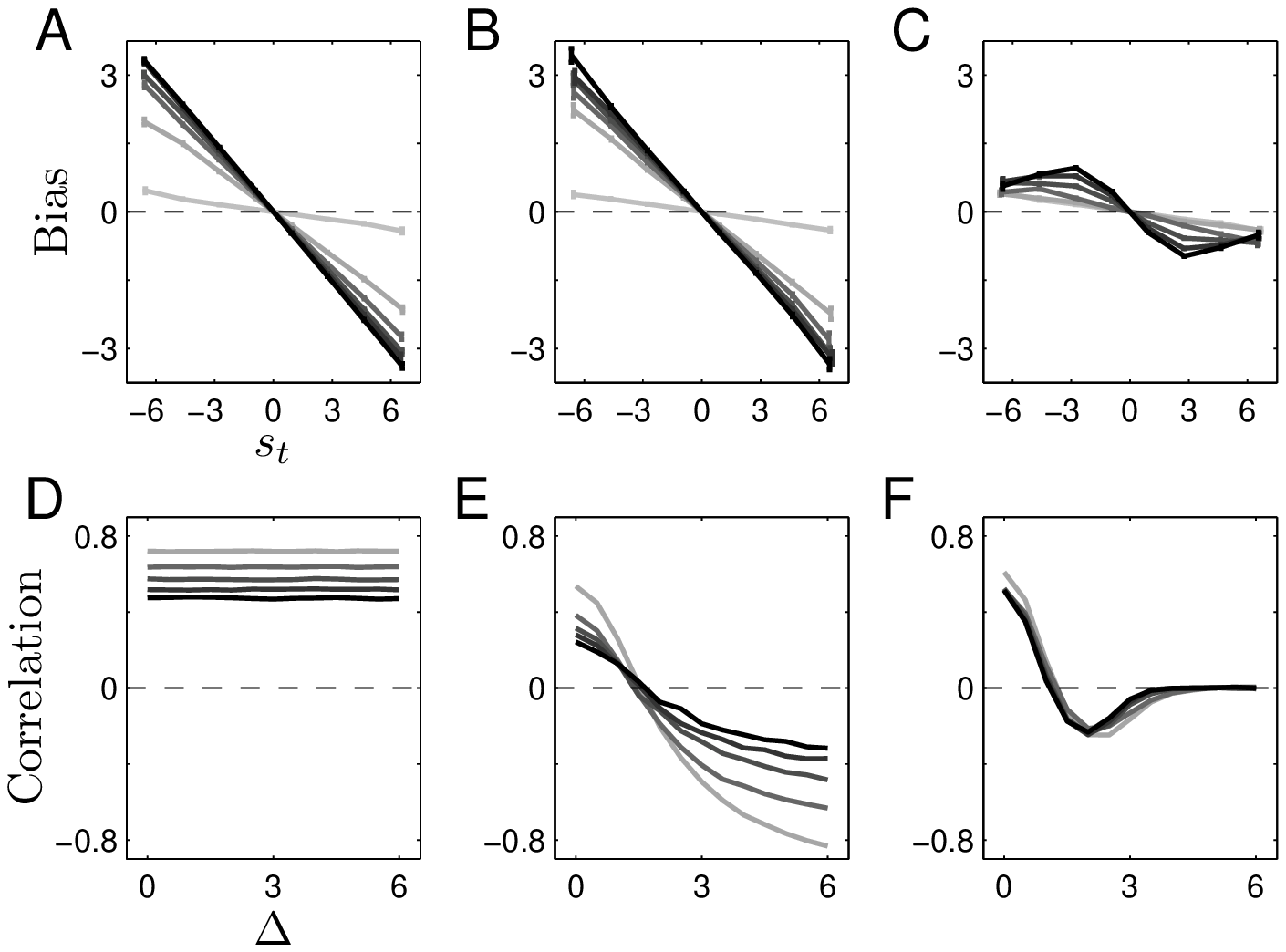} 
\caption{\textbf{Biases (A-C) and dependencies (D-F) predicted by the three models for $q(\mathbf{s})$ considered in the paper.} Biases (A-C) are computed by dividing the target stimulus value $s_t$ into 8 bins and measuring the mean and the SEM of the estimation errors, $\hat{s}_t - s_t$, for each bin. The experimenter's prior $p(\mathbf{s})$ is taken to be an uncorrelated multivariate Gaussian. Dependencies (D-F) are computed by presenting a particular set of stimuli consisting of a variable stimulus $s_1$ and $N-1$ zeros over $10^5$ simulated trials and measuring the correlation coefficient between the estimates of $s_1$ and $s_2$ over all trials (note that $s_2$ is always zero). $\Delta$ denotes the absolute difference between $s_1$ and $s_2$ ($\Delta = |s_1-s_2|$). For biases, simulation results are shown for 6 different set sizes (from $N=1$ to $N=6$). For dependencies, simulation results are shown for 5 different set sizes (from $N=2$ to $N=6$, as correlation does not have any meaning for $N=1$). In each plot, darker lines correspond to larger set sizes. (A) and (D) show the results for a multivariate Gaussian $q(\mathbf{s})$ with uniform correlations ($\rho_q = 0.96$); (B) and (E) show the results for the case of multivariate Gaussian $q(\mathbf{s})$ with random positive correlations ($\gamma = 0.15$); (C) and (F) show the results for the mixture of Gaussians prior ($k=1$).}
\label{bias_dependence_fig}
\end{figure}

Figure~\ref{bias_dependence_fig} shows the biases and dependencies predicted by the three models for $q(\mathbf{s})$ considered in the paper. The simulations shown here use a set size independent noise distribution $p(\mathbf{x}|\mathbf{s})$ (see the figure caption for more details about the simulations). For all models, biases increase with set size (Figure~\ref{bias_dependence_fig}A-C), consistent with the results reported in a previous study (Wilken \& Ma, 2004). Correlations between the estimates of different stimuli, on the other hand, decrease with set size (Figure~\ref{bias_dependence_fig}D-F). The large negative correlations observed in the multivariate Gaussian model with random positive correlations (and to a lesser degree in the mixture of Gaussians model) are due to the fact that in some trials the dependence between two stimuli under $q(\mathbf{s})$ is small and therefore produces a small bias in the estimates toward the prior mean, whereas in other trials the dependence between two stimuli under $q(\mathbf{s})$ is large and produces a larger bias toward the prior mean. Over many trials, this variability in biases generates an overall negative correlation.  

We note that the magnitude of biases shown in Figure~\ref{bias_dependence_fig}A-C is not inconsistent with those reported in Wilken and Ma (2004) [see Figure~8 in Wilken and Ma (2004)]. Orhan and Jacobs (2013) also reported large biases in a VSTM recall experiment for spatial location. However, not all studies find consistently large biases in subjects' responses. For example, Orhan and Jacobs (2013) found substantially smaller biases in a change detection task for spatial location. In addition, the constant correlations predicted by the Gaussian model with uniform correlations (Figure~\ref{bias_dependence_fig}D) and the negative correlations predicted by the Gaussian model with random positive correlations (Figure~\ref{bias_dependence_fig}E) and the mixture of Gaussians model (Figure~\ref{bias_dependence_fig}F) appear to be qualitatively inconsistent with the correlations reported in Orhan and Jacobs (2013).

Attributing part of the observed performance limitations (e.g., declines in precision with set size) to a general resource limitation reduces the amount of dependence in the subject's prior that would be needed to explain the empirical performance limitations (see the next section), which might, in turn, lead to smaller biases and dependencies in the modeled responses. Similarly, as shown in the main text (see the section ``Possible changes in the noise distribution due to adaptation to natural stimulus statistics''), stimulus-dependent noise distributions adapted to the subject's prior can generate biases away from the prior (see Stocker \& Simoncelli, NIPS '05, for a detailed explanation of this effect), which further counteract the bias toward the prior induced by the correlated priors.

\subsection*{Text S5. Weaker hypothesis: Model mismatch and capacity limitations together account for performance limitations in VSTM}
\label{weak_hyp} 

Here, we investigate the consequences of combining the effects of model mismatch and a general resource limitation on performance limitations in VSTM. To this end, we repeat the simulations reported in the main text, but now using a set size dependent noise distribution $p(\mathbf{x}|\mathbf{s})$. Specifically, we assume that the noise distribution is multivariate Gaussian with an isotropic, diagonal covariance matrix $\lambda^2 I$ where $I$ is the $N \times N$ identity matrix and $\lambda^2 = \lambda_{min}^2 N$ ($\lambda_{min}^2$ is the minimum achievable variance; we take $\lambda_{min}^2=0.25$) consistent with the theoretical arguments discussed in the main text about the increase in noise variance with set size due to a general resource limitation. Other details of the simulations reported here are the same as the corresponding simulations in the main text.

\subsubsection*{Memory precision-set size relationship}
\label{weak_setsize_sec}
\begin{figure}
\centering
\includegraphics[scale=1]{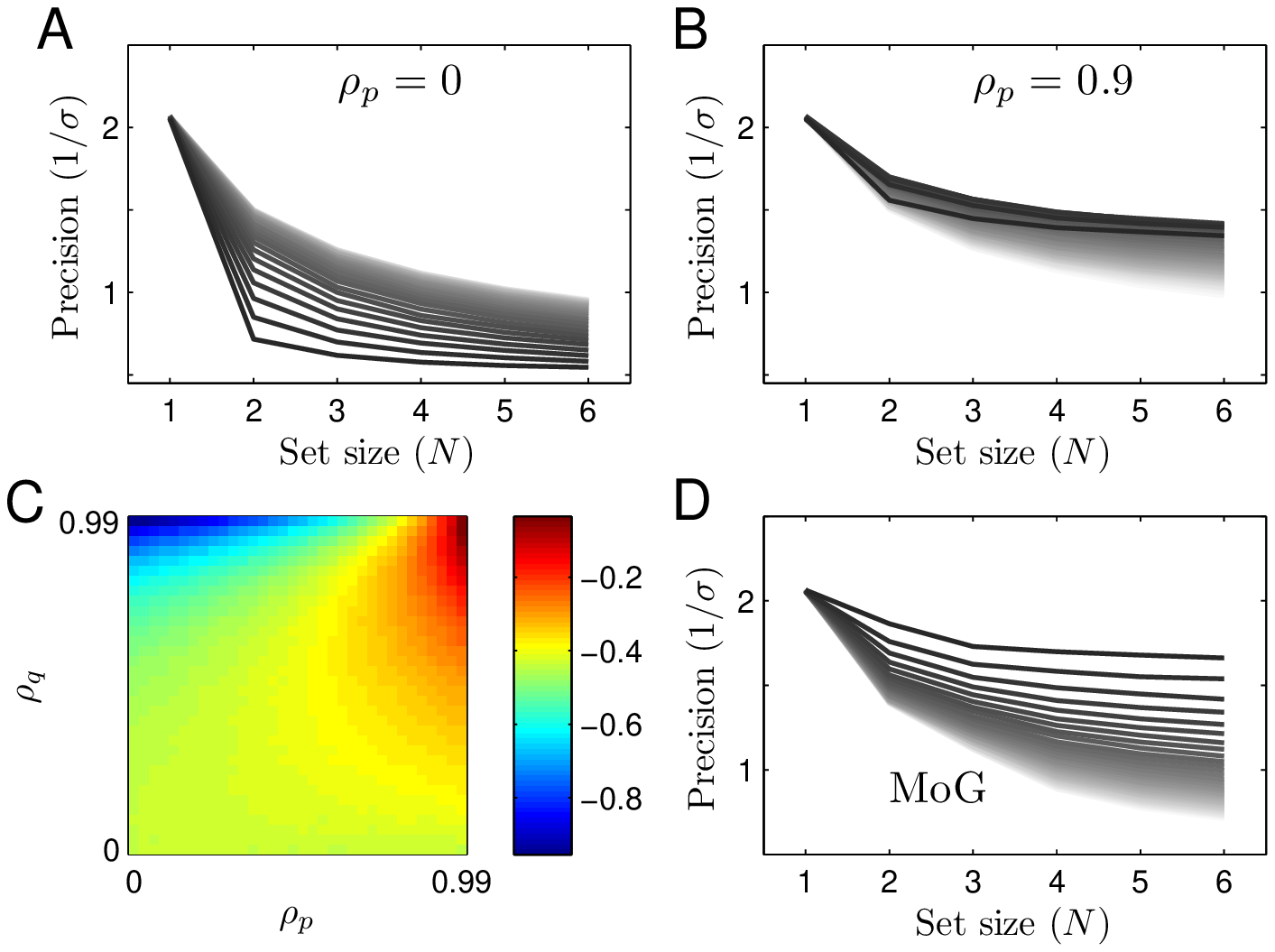} 
\caption{\textbf{Recall precision as a function of set size (when a resource limitation is combined with model mismatch).} (A) Recall precision as a function of set size when $p(\mathbf{s})$ is an uncorrelated Gaussian ($\rho_p=0$) and $q(\mathbf{s})$ is a multivariate Gaussian with uniform non-negative correlations $\rho_q$. Precision-set size curves are shown for 34 different values of $\rho_q$ from $0$ to $0.99$ in increments of 0.03. Lighter colors represent lower $\rho_q$ values. (B) Similar to A, but $p(\mathbf{s})$ is a multivariate Gaussian with uniform correlations with $\rho_p=0.9$. (C) Exponents $\xi$ of power-law fits to precision-set size curves for each pair of ($\rho_p$, $\rho_q$) values. (D) Results for the mixture of Gaussians model. $p(\mathbf{s})$ is a multivariate Gaussian with uniform non-negative correlations, and $q(\mathbf{s})$ is a mixture of Gaussians with the same parameter values used in previous simulations ($k=1$). $\rho_p$ is varied from $0$ to $0.99$ in increments of 0.03. Each curve represents the precision-set size curve for a different $\rho_p$ value. Lighter colors represent lower $\rho_p$ values.}
\label{correlated_pq_d_fig}
\end{figure}

How is the precision--set size relationship discussed in the main text affected if the noise variance increases with set size? Figure~\ref{correlated_pq_d_fig}A shows the recall precision as a function of set size when the experimenter's prior is an uncorrelated multivariate Gaussian and the subject's prior is a multivariate Gaussian with uniform correlations, where the correlation coefficient $\rho_q$ is varied between 0 and 0.99 in increments of 0.03. Figure~\ref{correlated_pq_d_fig}A should be compared to Figure~3A in the main text. Predictably, the increase in noise variance with set size leads to steeper declines in recall precision. Without any model mismatch (i.e., $\rho_p=\rho_q=0$), recall precision decreases as a power-law function of $N$ with an exponent of $-0.44$ (the slightly greater than $-0.5$ exponent is due to the use of a relatively informative prior in our simulations). Model mismatch decreases this exponent further, with larger mismatches leading to smaller exponents. A correlation coefficient of $\rho_q=0.81$ leads to $\xi = -0.61$ ($R^2=0.99$) and $\rho_q=0.84$ produces $\xi = -0.65$ ($R^2=0.99$). This result suggests a possible explanation of the steeper than predicted declines in precision observed in VSTM studies. Recall that in the presence of a fixed resource limitation, theoretical arguments predict the noise variance to scale as $N$, which in turn corresponds to a precision (inverse standard deviation) that scales as $N^{-0.5}$ (assuming a relatively non-informative prior). But, as discussed in the main text, most VSTM studies report power-law relationships between recall precision and set size with exponents ranging from $-0.60$ to $-0.75$. A possible explanation of this discrepancy is that the additional decline in precision, not explained by the increase in noise variance, might be caused by model mismatch.  

Introducing correlations in the experimenter's prior ($\rho_p=0.9$) leads to overall increases in recall precision and consequent increases in $\xi$ (Figure~\ref{correlated_pq_d_fig}B). For instance, $\rho_q=0$ yields an exponent of $\xi=-0.44$ and $\rho_q=0.9$ yields $\xi=-0.22$. However, in no cases do we find an inverse set size effect observed in the previous section where the noise distribution was set size independent. This is because the dependencies in the experimenter's prior are not strong enough to overcome the effect of increasing noise with set size. Figure~\ref{correlated_pq_d_fig}C shows the power-law exponents $\xi$ for all pairs of $(\rho_p,\rho_q)$ values tested. Figure~\ref{correlated_pq_d_fig}D shows the simulation results for the mixture of Gaussians prior for $q(\mathbf{s})$ with $k=1$. For the mixture of Gaussians prior, simulations with $\rho_p=0$ yield $\xi = -0.60$ ($R^2=0.99$) and simulations with $\rho_p = 0.9$ yield $\xi = -0.25$ ($R^2=0.97$).

For the remaining simulations reported here in the Supporting Information, we use parameter values for the models that, when combined with the set size dependent noise distribution, produce power-law exponents for the precision--set size relationship that are in the empirically observed range of $-0.60$ to $-0.75$. In particular, we use $\rho_q=0.81$ for the multivariate Gaussian model with uniform correlations, $\gamma=0.3$ for the multivariate Gaussian model with random positive correlations and $k=1$ for the mixture of Gaussians model.

\subsubsection*{Variability in memory precision}
\label{weak_variability_sec}
\begin{figure}
\centering
\includegraphics[scale=1]{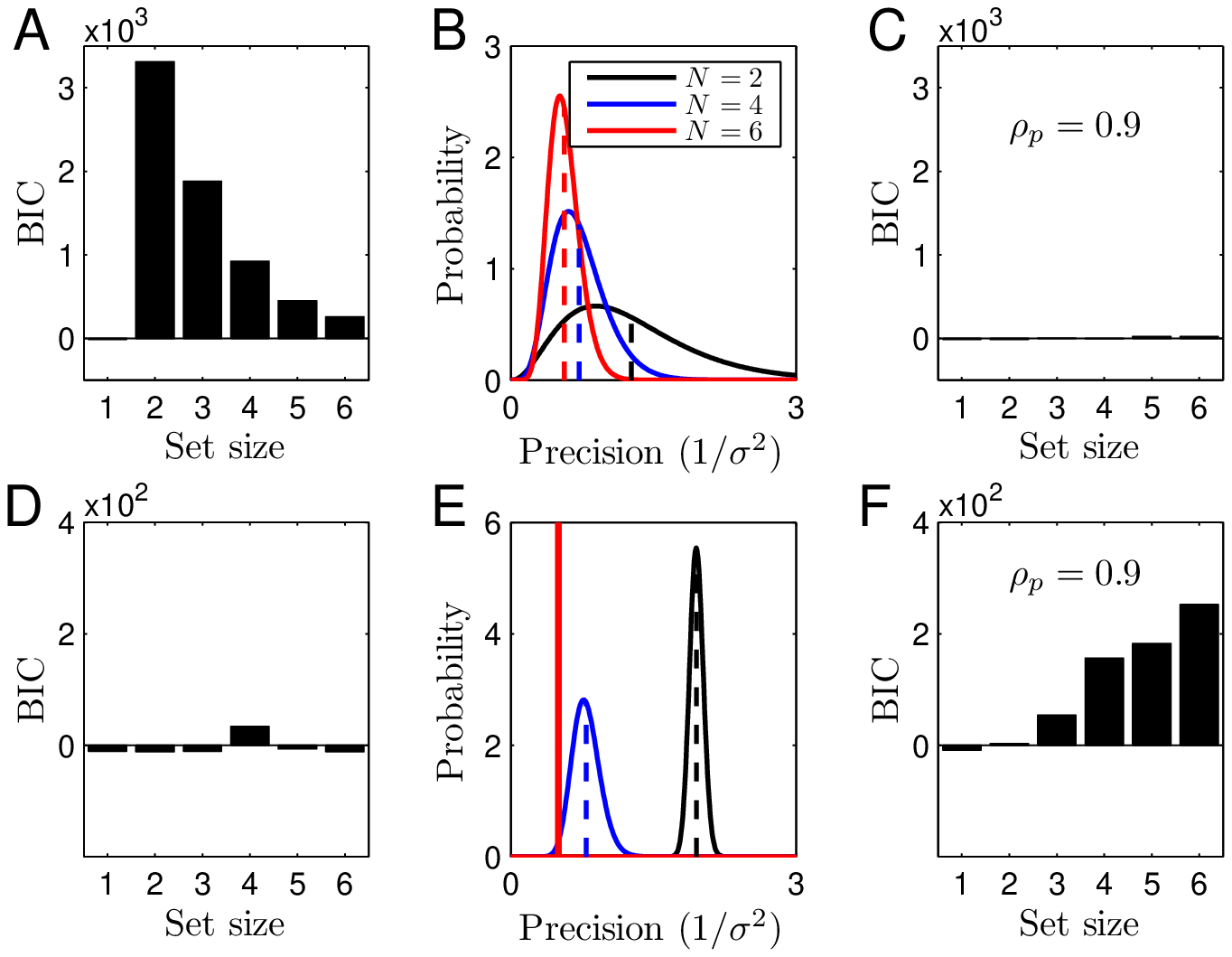} 
\caption{\textbf{Variability in memory precision when the noise variance increases with set size.} (A)-(C) show the results for a multivariate Gaussian $q(\mathbf{s})$ with random positive correlations ($\gamma$ = 0.3), (D)-(F) show the results for the mixture of Gaussians model ($k=1$). The formats of the plots are identical to those in Figure~5 in the main text.}
\label{variability_d_fig}
\end{figure}

Figure~\ref{variability_d_fig} shows the simulation results for variability in memory precision when the noise variance increases with set size. For the multivariate Gaussian $q(\mathbf{s})$ with random positive correlations ($\gamma=0.3$), the results are qualitatively very similar to the results obtained with a set size independent noise distribution. When the experimenter's prior is an uncorrelated Gaussian, there is significant variability in precision for all set sizes except $N=1$ (Figure~\ref{variability_d_fig}A-B). This variability is significantly reduced when the experimenter's prior displays strong correlations ($\rho_p=0.9$; Figure~\ref{variability_d_fig}C).

For the mixture of Gaussians prior, there is significant variability in memory precision only for $N=4$ when the experimenter's prior is an uncorrelated Gaussian (Figure~\ref{variability_d_fig}D-E), and for all set sizes except $N=1$ when the experimenter's prior has strong correlations (Figure~\ref{variability_d_fig}F). In contrast to the results reported in the main text with a set-size independent noise distribution, overall variability in memory precision is significantly larger for $\rho_p=0.9$ than for $\rho_p=0$. This is due to the relatively larger noise variances used in the simulations reported here. Through simulations (not shown), we confirmed that using a larger but set size independent noise variance produces results that are qualitatively quite similar to those shown in Figure~\ref{variability_d_fig}D-F.

We emphasize that, for any given set size, it is still the properties of the subject's prior and its interaction with the experimenter's prior that primarily generate variability in memory precision. The set size dependence of the noise distribution by itself cannot explain this variability. The magnitude of the noise variance only modifies the amount of variability in memory precision.

\subsubsection*{Different set-size dependencies for initial encoding rate and asymptotic precision}
\label{weak_encodingrate_sec}
\begin{figure}
\centering
\includegraphics[scale=1]{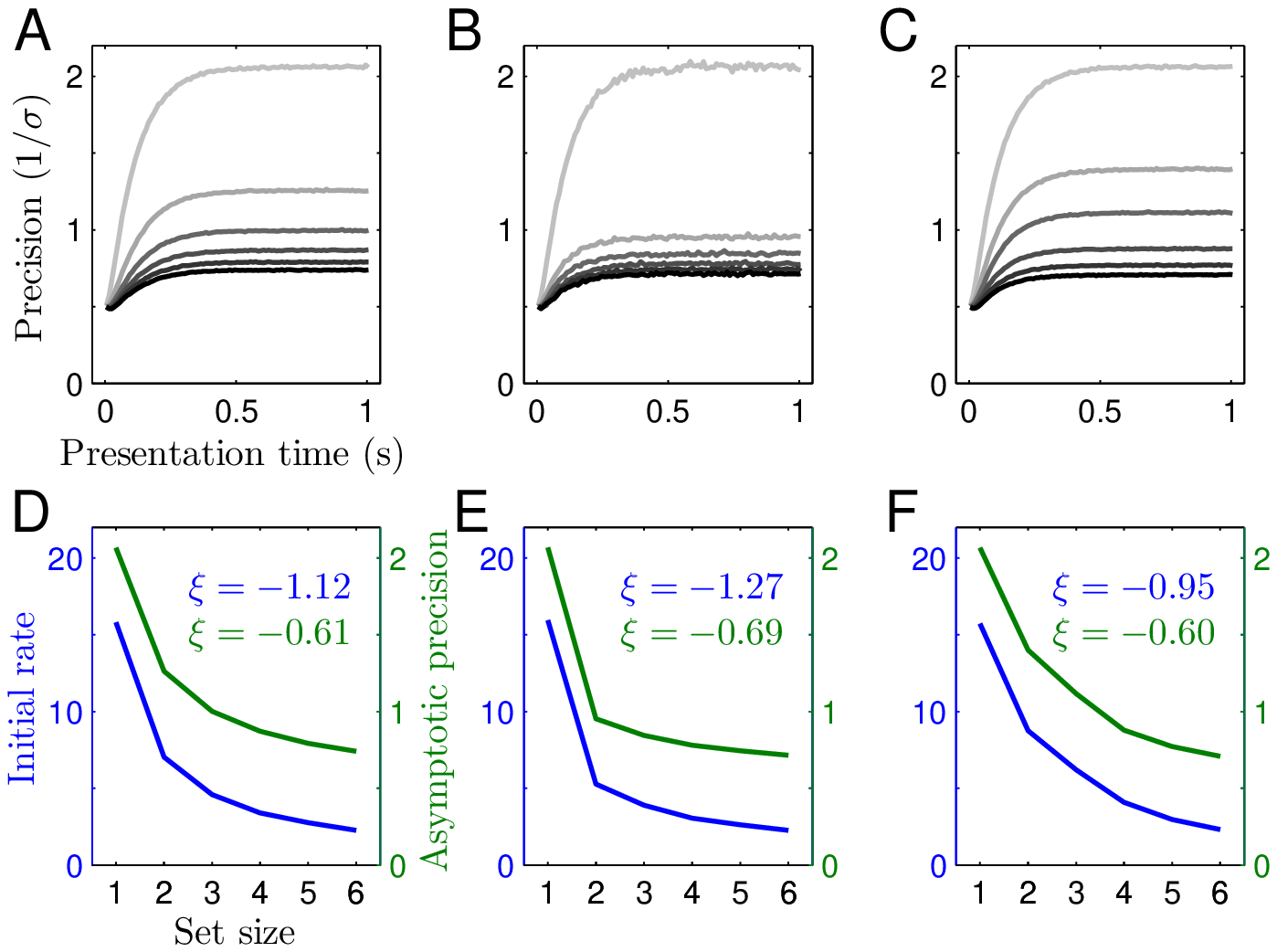} 
\caption{\textbf{Time course of recall precision when the noise variance increases with set size and the experimenter's prior $p(\mathbf{s})$ is an uncorrelated Gaussian.} The formats of the plots are identical to those in Figure~6 in the main text. (A) and (D) show the results for a multivariate Gaussian $q(\mathbf{s})$ with uniform correlations ($\rho_q=0.81$). (B) and (E) show the results for a multivariate Gaussian $q(\mathbf{s})$ with random positive correlations ($\gamma=0.3$). (C) and (F) show the results for the mixture of Gaussians prior ($k=1$).}
\label{encodingrate_d_fig}
\end{figure}

\begin{figure}
\centering
\includegraphics[scale=1]{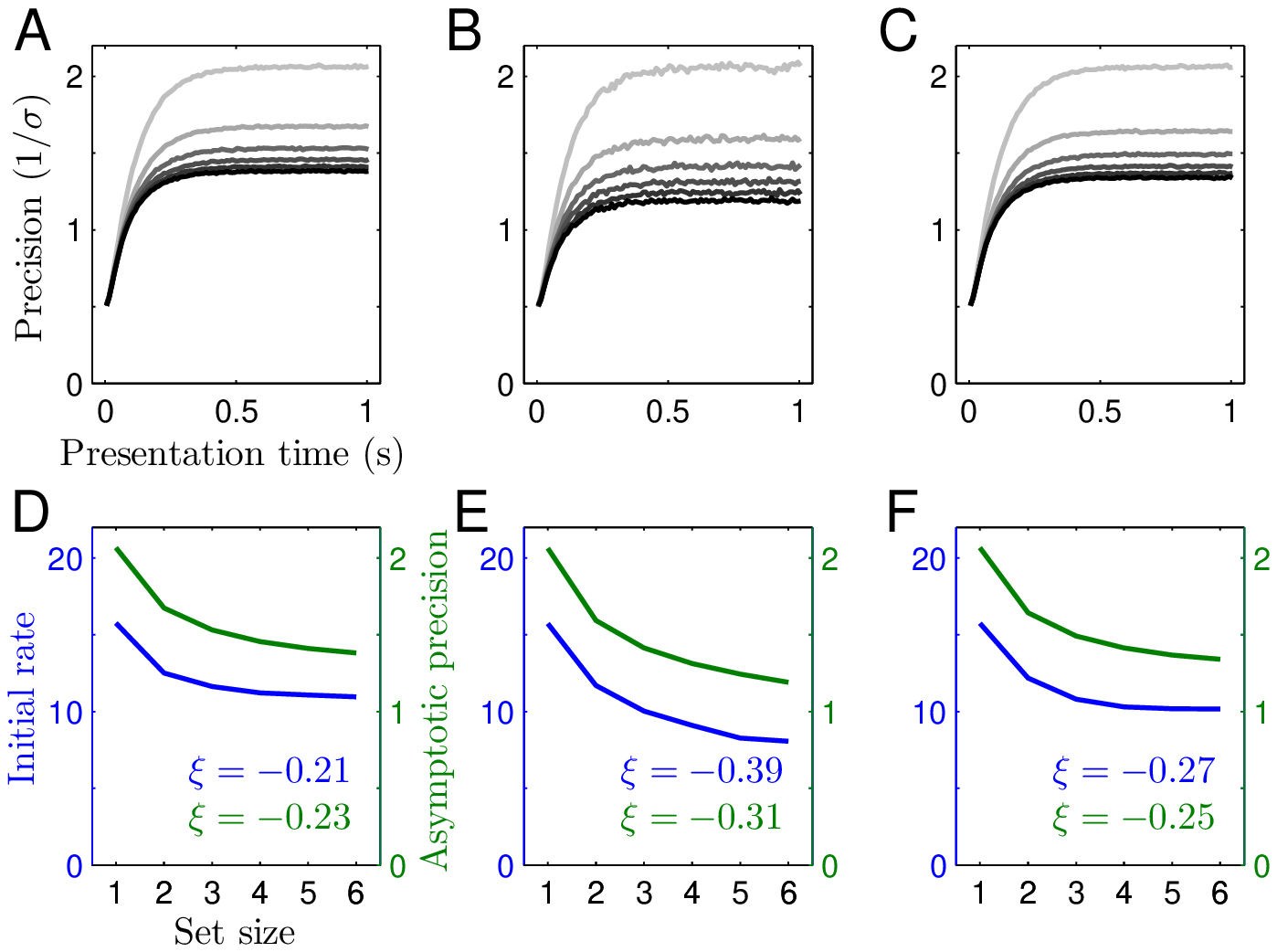} 
\caption{\textbf{Time course of recall precision when the noise variance increases with set size and the experimenter's prior $p(\mathbf{s})$ is a multivariate Gaussian with uniform correlations ($\rho_p=0.9$).} The formats of the plots are identical to those in Figure~\ref{encodingrate_d_fig}.}
\label{encodingrate_d_corr_fig}
\end{figure}

Figures~\ref{encodingrate_d_fig} and \ref{encodingrate_d_corr_fig} show the simulation results on the time course of recall precision when the noise variance increases with set size. Figure~\ref{encodingrate_d_fig} shows the results for the case where the experimenter's prior $p(\mathbf{s})$ is an uncorrelated Gaussian and Figure~\ref{encodingrate_d_corr_fig} shows the results for the case where $p(\mathbf{s})$ is a multivariate Gaussian with uniform correlations (with $\rho_p=0.9$).  

In these simulations, we assume that the precision of the noise distribution changes with presentation time according to Equation~4 in the main text. In addition, however, we also assume that $P_{max}$ changes with set size according to $P_{max} = \kappa N^{-0.5}$, where $\kappa$ is the maximum achievable precision (obtained for $N=1$). We set $\kappa=2$. Importantly, however, we assume that $\tau$ in Equation~4 (in the main text) does not depend on the set size (we set $\tau=0.1$). 

The results are qualitatively similar to the results obtained with a set size independent noise distribution. Different set-size dependencies of the initial encoding rate and the asymptotic precision are successfully captured by all three models (Figure~\ref{encodingrate_d_fig}). The models also yield power-law exponents that are similar to the values reported in Bays et al. (2011): $-1.01 \pm 0.14$ for the initial encoding rate and $-0.60 \pm 0.12$ for the asymptotic precision. 

Introducing strong correlations in the experimenter's prior ($\rho_p=0.9$) also produces qualitatively similar effects to those found in the main text with a set size independent noise distribution. In particular, correlations in the experimenter's prior reduce the difference between the power-law exponents for the initial encoding rate and the asymptotic precision (Figure~\ref{encodingrate_d_corr_fig}). Unlike in the previous section, however, only the multivariate Gaussian $q(\mathbf{s})$ with uniform correlations displays a crossover between the exponents for the initial encoding rate and asymptotic precision (Figure~\ref{encodingrate_d_corr_fig}D). With the particular parameter values used in these simulations, the other two models predict only a reduction in the difference between the exponents without a change in their ordinal relationship. 

\subsection*{Text S6. Set-size dependence of the analytic estimates of initial encoding rate and asymptotic precision}
\label{analytic_encodingrate_precision} 
For the Gaussian model with uniform correlations, using a time-dependent noise variance $\sigma^2(t)$ in the analytic expression for the precision of the error distribution (Equations~\ref{part_1}-\ref{part_2}), and taking the derivative of the resulting expression with respect to $t$ gives us an analytic expression of the encoding rate as function of time $t$ (intuitively, how fast information is encoded into VSTM at time $t$). In the main text, we assume $1/\sigma(t) = P_{max} (1 - \exp(-t/\tau))$ where $P_{max}$ and $\tau$ are set-size independent. Because the analytic expression for the precision of the error distribution is a very complicated function of $1/\sigma$, we computed the derivative of the precision of the error distribution (Equations~\ref{part_1}-\ref{part_2}) with respect to $1/\sigma(t)$ using Matlab's Symbolic Math toolbox. The time derivative was then obtained with the chain rule. 

\begin{figure}
\centering
\includegraphics[scale=1]{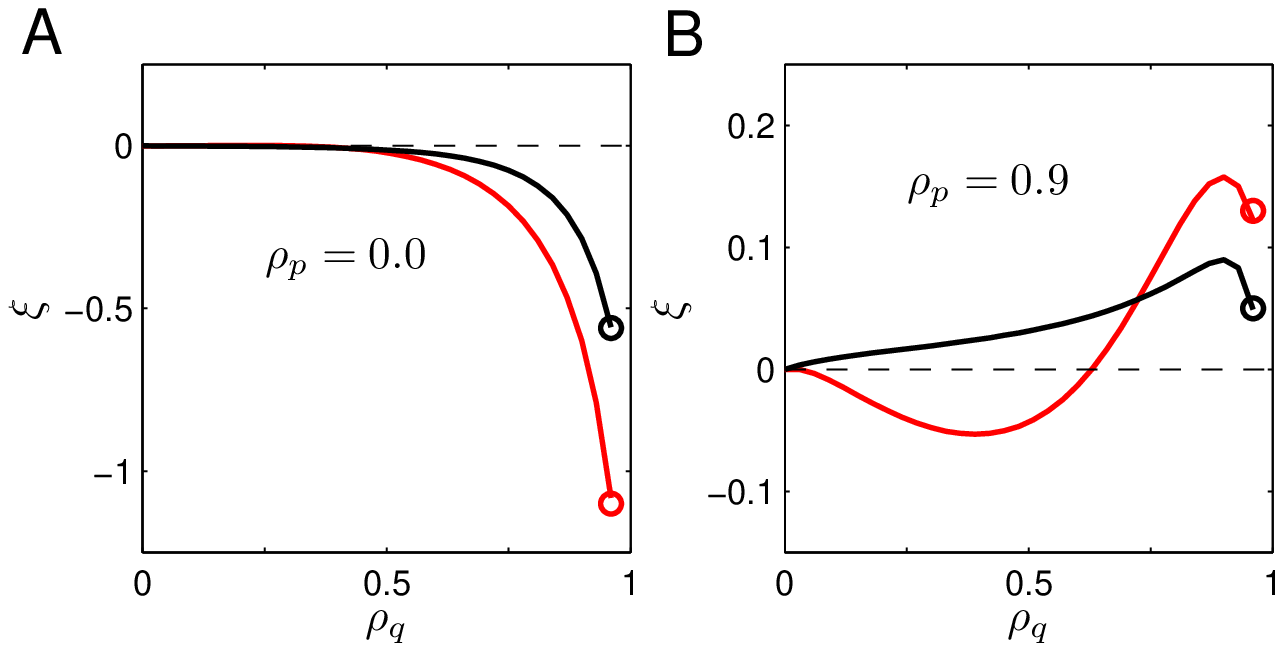} 
\caption{\textbf{Set-size dependencies of the analytic estimates of initial encoding rate (red) and asymptotic precision (black) in the Gaussian model with uniform correlations.} Initial encoding rates were estimated from the time derivative (at time 40 ms) of the analytic expression for precision. Set sizes ranged from 1 to 6. Other parameters were the same as in the simulations reported in the main text (Figure 6A and D). (A) Exponents of power-law fits, $\xi$, as a function of the correlations in the subject's prior, $\rho_q$, when the experimenter's prior is an uncorrelated Gaussian ($\rho_p = 0$). (B) Exponents of power-law fits as a function of $\rho_q$ when the experimenter's prior is a correlated Gaussian ($\rho_p = 0.9$). Note the crossover point where the red line intersects the black line. Circles indicate the simulation results reported in the main text (Figure 6D and 8D).}
\label{encodingrate_precision_all}
\end{figure}

Figure~\ref{encodingrate_precision_all} shows the set-size dependencies of the analytic estimates of initial encoding rate (red) and asymptotic precision (black) as a function of the correlation coefficient of the subject's prior. Analytic estimates of the initial encoding rate were obtained from time derivatives at time 40 ms to provide a good fit to simulation results. This is legitimate because the initial encoding rates estimated in the simulations were obtained only indirectly by fitting negative exponential functions to the time-course of precision. Analytic estimates obtained from derivatives at earlier or later times give qualitatively similar, but quantitatively different results.

\end{document}